\documentclass[acmsmall]{acmart}

\usepackage[utf8]{inputenc}
\usepackage{bm}
\usepackage[ruled,vlined,linesnumbered]{algorithm2e}
\usepackage{algorithmic}
\usepackage{booktabs}
\usepackage{tabularx, hhline}
\usepackage{diagbox}
\usepackage{bm}
\usepackage{subcaption}

\usepackage[english]{babel}
\ifodd 0
    \newcommand\rev[1]{{\color{red}#1}}

    \newcommand{\com}[2]{\textbf{\color{blue} (COMMENT from [#1]: #2)}}
\else

    \newcommand\rev[1]{{#1}}
    
    \newcommand{\com}[2]{}
\fi

\begin{document}

\title[Optimal Competitive Online Peak Minimization]{Optimal Online Peak Minimization Using Energy Storage}

\author{Yanfang MO, Qiulin LIN, Minghua CHEN, and Si-Zhao Joe Qin}
\affiliation{%
  \institution{City University of Hong Kong}
  \city{Hong Kong}
  \country{China}}

\renewcommand{\shortauthors}{Submission ID: \#XX}

\begin{abstract}
The significant presence of demand charges in electric bills motivates large-load customers to utilize energy storage to reduce the peak procurement from the grid. We herein study the problem of energy storage allocation for peak minimization, under the online setting where irrevocable decisions are sequentially made without knowing future demands. The problem is uniquely challenging due to (i) the coupling of online decisions across time imposed by the inventory constraints and (ii) the noncumulative nature of the peak procurement. We apply the CR-Pursuit framework and address the challenges unique to our minimization problem to design an online algorithm achieving the optimal competitive ratio (CR) among all online algorithms. We show that the optimal CR can be computed in polynomial time by solving a linear number of linear-fractional problems. More importantly, we generalize our approach to develop an \emph{anytime-optimal} online algorithm that achieves the best possible CR at any epoch, given the inputs and online decisions so far. The algorithm retains the optimal worst-case performance and attains adaptive average-case performance. Trace-driven simulations show that our algorithm can decrease the peak demand by an extra 19\% compared to baseline alternatives under typical settings.

\end{abstract}

\maketitle

\section{Introduction}
\begin{figure}[t]
    \centering \centerline{\includegraphics[width=0.7\columnwidth]{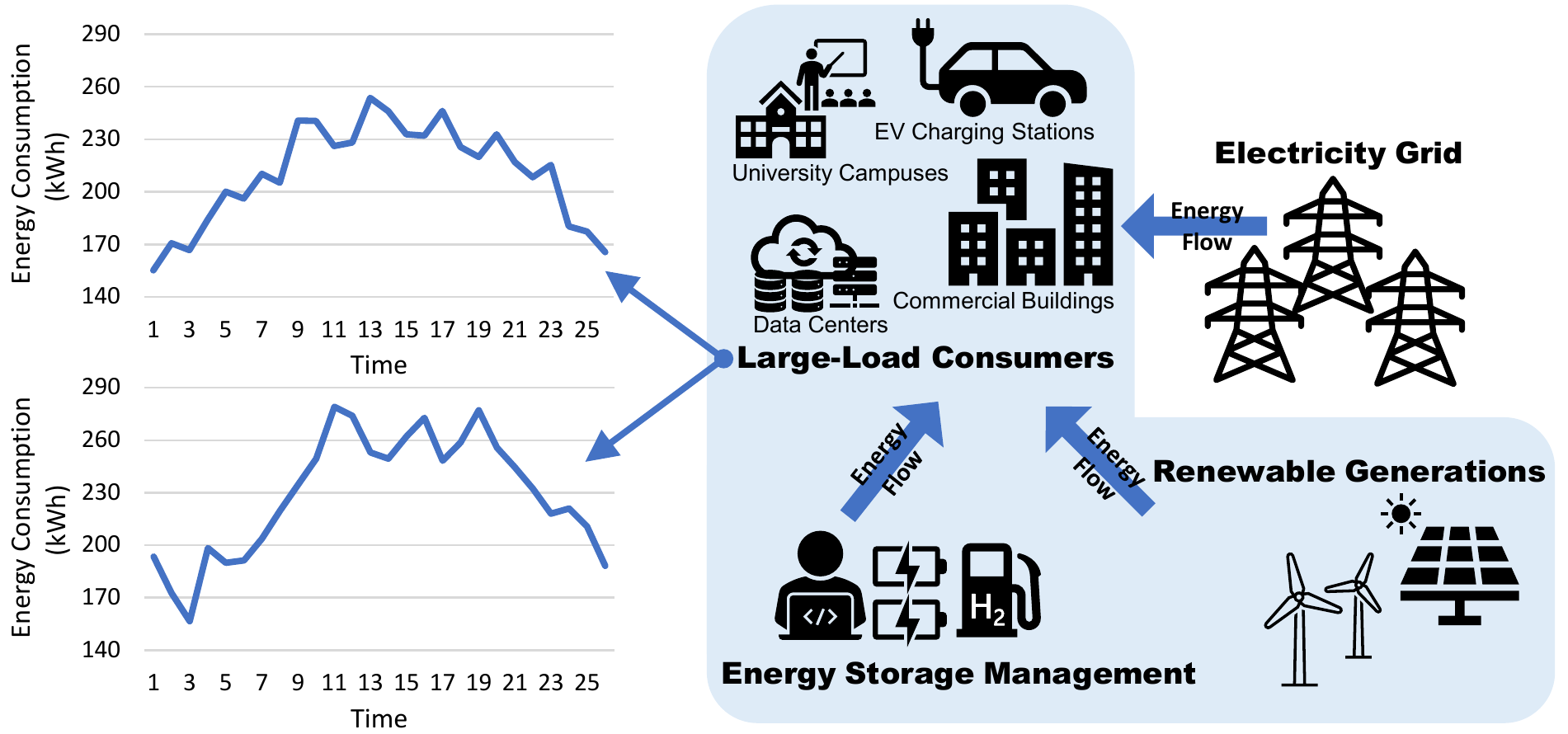}}
    \caption{Real-time energy consumption in two on-peak periods~(left) and an
    illustrative scenario of utilizing energy storage systems to reduce peak
    usages for large-load consumers with behind-the-meter renewable generations~(right).
    }%
    \label{fig: ExampLoad}
\end{figure}\noindent


Demand side management~(DSM) promotes desirable changes in load curves to ease the supply side burden~\cite{gellings1985concept}. A popular DSM mechanism is that utilities use pricing schemes to motivate customers to improve their consumption patterns. Many pricing schemes are designed
to reflect the cost of energy over time, e.g., time-of-use~(TOU) and real-time pricing~(RTP)~\cite{newsham2010effect,sgouras2017quantitative}.
These schemes use price differences to incentivize load shifting from
on-peak hours, when the supply is scarce or expensive, to off-peak
hours with abundant or cheap supply.

Meanwhile, utilities introduce peak-demand charges to encourage large-load customers, e.g., data centers and shopping
malls, to flatten their load curves, especially during the on-peak
hours~\cite{newsham2010effect,liu2013data}. The peak-demand charge is the maximum rate at which a customer pays for the amount of power (demand) \rev{during a prespecified period}\footnote{\rev{The demand charge is typically settled on a monthly basis~(\$/kW/month). Utilities also released tariffs with daily demand charges (\$/kW/day)~\cite{Daily_DM,vaghefi2014hybrid}.}}. The demand is often measured in~$15$-minute or~$30$-minute intervals, with the highest measured value setting the ``peak''
for which the customer is billed. The unit of measure is typically the kilowatt (kW). \rev{Notably, the peak-demand charge can be~$80\%$--$90\%$ of the total electricity bill of a large-load user like a charging station~\cite{lee2020pricing} or data center~\cite{xu2014reducing}.}

It is thus financially attractive to invest in technologies
for reducing the peak electricity usage~\cite{xu2014reducing,DemandChargeTBR,shi2017using}.
See Fig.~\ref{fig: ExampLoad} for the illustration of an increasingly
popular scenario, where large-load customers, e.g., the commercial, the industrial,
and the municipal, utilize energy storage to reduce the peak power procurement
from the grid, in the presence of self-owned renewable generations like wind turbines and solar panels. 
\rev{We herein focus on the discharging of energy storage during on-peak periods; thus, the proposed algorithms directly apply to diverse energy storage systems that may not even admit charging by the grid or the consumer-owned renewable like fuel cells~\cite{BloomEnergy} and shared storage~\cite{kalathil2017sharing}\footnote{\rev{The focus on the discharging-only setting is of interest for
the following reasons. First, it is a simple yet popular strategy
to charge batteries during off-peak hours and discharge in
on-peak periods to reduce the peak usage~\cite{kalathil2017sharing}. Second, frequent charging/discharging may greatly shorten the service time of batteries~\cite{he2015optimal,xu2016modeling,foggo2017improved}. Also, certain storage systems, e.g., fuel cells~\cite{BloomEnergy} and
shared storage~\cite{kalathil2017sharing}, do not allow charging by the grid or the renewable during on-peak hours. The high self-discharge rate makes charging flywheels in on-peak hours uneconomical~\cite{olabi2021critical}. Finally, solutions
for the discharging setting serve as benchmarks for those designed
for the charging/discharging setting.}}.} Customers' peak-minimizing endeavors also benefit the power system because flattened load curves reduce the risk of overloading transmission lines or transformers, the need of calling expensive and carbon-intensive reserve generators, and the pressure of managing curtailable electricity services~\cite{oren1992design} to serve surging loads.

However, it is algorithmically non-trivial to realize such a win-win
benefit in practice. It is hard to forecast the fluctuating
loads, especially peak usages for individual customers. Even worse,
the user-owned (behind-the-meter) intermittent renewable generations
aggravate the unpredictability of net demands to be satisfied
\cite{jacobsen2012curtailment}, leading to online (peak) optimization
with little or no future information\footnote{There are many results under the clairvoyant/stochastic setting
where the complete or statistical information of loads and renewable generations
is available \cite{alharbi2017stochastic,qin2015online,bertsimas2011theory,xiang2015robust,li2020real}. 
While achieving strong performance, they may not
be practical due to the difficulty of estimating loads~\cite{wang2019probabilistic} and the renewable~\cite{croonenbroeck2019renewable}.}. While online optimization is a well-studied topic, versions with
inventory constraints and peak charges are known to be difficult. The
coupling of online decisions across time imposed by the inventory
constraints and the noncumulative nature of the peak usage make the
problem uniquely challenging and little understanding is available
in the literature; see a discussion on the related work in Section~\ref{Sec_RelatedWork}.

Motivated by these observations, we study the problem
of online peak minimization under inventory constraints\footnote{There are also researches in maximizing the peak-demand reduction under the online setting, which is attractive to energy storage service providers whose profits are proportional to the peak-reduction of their customers. We discuss the difference between the two lines of endeavors in Sec.~\ref{Sec_RelatedWork} and Appendix~\ref{app_expdif}.}. We focus
on designing competitive algorithms under the online setting, where
the loads and renewable generations are revealed sequentially in time
but the algorithm has to make irrevocable decisions at current
epoch with little or no future information. We use \emph{competitive ratio} (CR) as the performance metric,
which is the worst-case ratio between the peak procurement attained
{by the online/real-time algorithm with little future information and that obtained by an optimal offline/clairvoyant
solution with complete knowledge of future inputs~\cite{borodin2005online,buchbinder2016unified}.} CR takes a value at least one and
we prefer online algorithms with smaller CRs, indicating that the online algorithms have closer performance to the optimal one in hindsight. In this work, we make the following contributions.

$\mbox{\ensuremath{\rhd}}$ After formulating the problem of peak
minimization under inventory constraints in Section~\ref{Sec_ProbForm},
we design an optimal online algorithm with the smallest CR among all deterministic and randomized algorithms for the problem in Section~\ref{Sec_OnlineAlg}.
We show the best CR can be computed in polynomial time, by solving
a linear number of linear-fractional programs. The best CR also captures the fundamental \emph{price of uncertainty} for the problem. To our
best knowledge, these are the first (and optimal) results on this
theoretically challenging online minimization problem that has wide application beyond the energy storage context. 

$\mbox{\ensuremath{\rhd}}$ In Section~\ref{Sec_Adaptive}, we generalize
our approach to develop an \emph{anytime-optimal} online algorithm.
It achieves the best possible CR at any epoch, given the inputs and
online decisions so far. The idea is to progressively prune the input
space based on the inputs observed so far and adjust the online
decisions for the best competitiveness with respect to the residual uncertainty. The algorithm retains the optimal worst-case performance
and achieves adaptive average-case performance.

\rev{
$\mbox{\ensuremath{\rhd}}$ In Section~\ref{Sec_Hybrid}, we discuss lowering the monthly demand charge based on our daily endeavor for reducing peak usages, electricity cost minimization, and other practical concerns.
}

$\mbox{\ensuremath{\rhd}}$ We carry out simulations based on real-world
traces to evaluate the empirical performance of our algorithms in Section~\ref{Sec_Simulation}. Under typical settings, our anytime-optimal algorithm improves the average daily peak reduction by over~$19\%$
as compared to baseline alternatives. It achieves up to~$77\%$ of the best possible daily peak reduction attained by the optimal offline algorithm. Note that the optimal offline algorithm is impractical since it is non-causal and assumes full knowledge of future demand and renewable generation. Our algorithm outperforms the receding horizon control~(RHC)~\cite{qin2003survey,ma2012demand} with a quarter of the operation period as the look-ahead window size. This highlights the usefulness of competitive algorithm design in decision making under uncertainty. Moreover, we show the numerical results on the monthly peak-demand minimization in Sec.~\ref{sim: monthly}, which further demonstrate the effectiveness of our online approach. 

\rev{We conclude the work in Section~\ref{Sec_Conclusion}. 
All proofs are deferred to the appendix.}

\section{Related Work}\label{Sec_RelatedWork}
\subsection{Electricity Cost Saving by Energy Storage}
Ref.~\cite{walawalkar2007economics} studied the economics of energy storage systems in New York state's electricity market by applications like energy arbitrage and regulation services. Ref.~\cite{urgaonkar2011optimal} used UPS units to reduce the time average electric utility bill for a single data center via a Lyapunov optimization approach. Ref.~\cite{guo2013electricity} further investigated the extended case with multiple data centers, under time-varying and location-varying electricity costs. A Lyapunov optimization-based online storage control was proposed in~\cite{shi2022lyapunov} to save costs for commercial buildings. Moreover, in~\cite{xi2014stochastic} and~\cite{roberts2019impact}, the authors respectively discussed the value of storage in a residential home and the impact of shared energy storage systems in apartment buildings. 

\subsection{Peak-Demand Charge}
Ref.~\cite{neufeld1987price} justified the usefulness of peak-demand charges for price discrimination given the competition among isolated industrial customers. Ref.~\cite{zakeri2014optimization} studied the demand charge incurred by the largest accumulated demand over several slots. Knowing the maximum gross demand, a client respectively used lossy and lossless buffering to minimize the maximum resource request in~\cite{bar2008reap} and~\cite{bar2008peak}. Ref.~\cite{shi2017using} used battery storage for peak shaving together with frequency regulation, based on day-ahead load prediction. Ref.~\cite{risbeck2020economic} applied the economic model predictive control for time-varying cost and demand charge optimization. Given a tariff with peak-demand charges, Ref.~\cite{zhao2017robust} studied the scheduling of EV charging jobs, while Ref.~\cite{lee2020pricing} studied the pricing of EV charging services. Our work specifically focuses on reducing the peak-demand charge by discharging limited energy storage in an online fashion.



\subsection{Competitive Analysis and Optimization under Uncertainty}

Online algorithm design with competitive analysis is widely used in smart grids, e.g., EV charging~\cite{tang2014online,zhao2017robust,yi2019balancing,sun2020competitive,lin2021minimizing} and economic dispatching~\cite{lin2012dynamic,badiei2015online,zhang2016peak,chau2016cost}. Online peak-demand minimization is uniquely hard by the noncumulative nature. Prior to this work, Refs.~\cite{zhao2017robust} and~\cite{zhang2016peak} respectively designed competitive online algorithms for EV charging and economic dispatching for microgrids, to minimize the peak consumption. Online optimization with inventory constraints challenges researchers by the coupling of online decisions across time. Classic instances include the one-way trading problem~\cite{el2001optimal} and variations of the knapsack problem~\cite{sun2020competitive}. Particularly, Ref.~\cite{lin2019competitive} concerned the online revenue maximization under inventory constraints and Ref.~\cite{yang2020online} studied the online linear optimization with inventory management constraints. Our study differs from them and complements the literature by considering both the noncumulative peak minimization objective function and inventory constraints. Furthermore, there is a useful algorithmic framework, called~\textsf{CR-Pursuit}. It leads to competitive algorithms with given CRs in~\cite{yi2019balancing,zhao2017robust,lin2021minimizing,lin2019competitive} and~\cite{hunsaker2003optimal}. {These pioneering results motivate us to design online peak-minimizing  algorithms parameterized by ``pursuing'' carefully designed CR at each decision-making epoch. }


There are also other popular approaches for optimization under uncertainty, e.g., robust optimization~\cite{bertsimas2011theory}, stochastic optimization~\cite{alharbi2017stochastic,qin2015online}, and model predictive control (MPC)~\cite{qin2003survey,risbeck2020economic}. Similarly to these methods, our approach adopts the worst-case analysis as robust optimization, concerns the average-case performance as stochastic optimization, and makes the decision at each round by virtue of a simple optimization problem as MPC. Meanwhile, as compared to competitive online optimization, robust optimization usually does not tackle sequential decisions. \rev{Stochastic optimization and MPC rely more heavily on reliable estimations of true distributions~\cite{zhao2018data} or accurate predictions of future values~\cite{qin2003survey}. However, such information is hard to obtain given volatile renewable generations~\cite{croonenbroeck2019renewable} and individual loads~\cite{wang2019probabilistic}. Moreover, MPC may not have a performance guarantee~\cite{ma2012demand,van2019comparison}.} Furthermore, these three approaches do not directly consider the fairness in performance evaluation under different inputs. In this paper, we develop online algorithms for minimizing peak demand using energy storage with the optimal worst-case competitive performance guarantee.


\section{Problem Formulation}\label{Sec_ProbForm}
We consider the scenario where large-load customers, e.g., shopping malls and data centers, utilize energy storage to lower their electricity bills. \rev{Demand charges have much higher rates than energy costs and can be over~$80\%$ of the bills~\cite{lee2020pricing,xu2014reducing,vaghefi2014hybrid,Daily_DM}. Usually, the monthly (\$/kW/month) or daily (\$/kW/day) demand charge is not assessed in all hours but a specified period of each day, overlapping with the on-peak period of a TOU tariff~\cite{sun2018break,Daily_DM,vaghefi2014hybrid}. Meanwhile, the energy consumption of many users typically cycles each day; moreover, flatter daily demand curves often mean more stability and certainty on energy prices, and less operating and environmental costs~\cite{luo2019optimal,ma2012demand,sun2018break}. Thus, we start with reducing the peak procurement from the grid over an on-peak period of a day. We divide it into~$T$ time slots, each of 15 or 30 minutes, matching the power measurement interval in practice. We convert the power demand (in kW) into energy consumption (in kWh) of a slot.}

\subsection{Mathematical Model}



\subsubsection*{Net Electricity Demand}

\rev{The large-load user may have user-owned (behind-the-meter) small-scale renewable generation (e.g, a commercial building with a solar roof), which has negligible generation cost and is insufficient to accommodate the gross demands in on-peak time slots. Note that we cannot charge energy storage systems based on fuel cells or shared storage by the on-site renewable. To cover these cases, we assume that the large load immediately consumes the self-owned renewable generation\footnote{\rev{We note that the selling price is usually no more than the purchase price for a user~\cite{fares2017impacts}; thus, the user had better consume the electricity generated by the renewable or from the energy storage rather than sell it back to the grid.}}.} Let~$d_{t}$ (in
kWh) be the residual demand not balanced by the local renewable
generation at time slot~$t\in[T]$. Then, we obtain the net demand profile~$\bm{d}\in\mathbb{R}^{T}$. We assume the minimum prior knowledge of the net demand, namely~$d_{t}\in[\underline{d},\overline{d}]$ for all~$t\in[T]$, where~$\underline{d}\in\mathbb{R^{+}}$ and~$\overline{d}\in\mathbb{R^{+}}$
are respectively the lower and upper bounds of the net demand in a
single time slot\footnote{The approach and analysis herein can be generalized to the case where each slot relates to a different pair of lower and upper bounds of net demand.}.
Note that we do not rely on any specific stochastic model of~$d_{t}$,~$t\in[T]$, which can be difficult to obtain because of the fluctuating individual loads and the volatile small-scale renewable generations.

\subsubsection*{Energy Storage}
For a battery energy storage system, it is a common practice to charge the storage during off-peak periods with lower TOU prices and discharge it during on-peak periods when the electricity is dear~\cite{kalathil2017sharing}. In this way, we expect to decrease the demand charge and exploit TOU energy arbitrage; moreover, we can reduce the degradation cost and better estimate the lifespan of batteries as a result of controlling the number of charge/discharge cycles and depth of discharge~\cite{he2015optimal,xu2016modeling,foggo2017improved}. On the other hand, if the energy storage system is based on fuel cells, e.g., Bloom Energy servers~\cite{hardman2015fuel}, or the storage shared by a sharing aggregator~\cite{kalathil2017sharing}, the user cannot charge the storage by the grid. These facts motivate us to focus on discharging limited energy storage in a considered on-peak period. Let~$c$ and~$\bar{\delta}$ (in kWh) respectively denote the available storage capacity and the maximum discharge amount per slot\footnote{\rev{Here the capacity~$c$ is defined as the product of the discharge efficiency ratio (in~$(0,1]$) and the maximum depth of discharge predetermined before the period.} The discharging efficiency defines the ratio of the energy received by the demand and that taken out of the system. Similarly, the discharge rate limit~$\bar{\delta}$ is factored in the consideration of the discharging efficiency.}. Customers may hybridize storage technologies with complementary characteristics of power and energy density for desired~$c$ and~$\bar{\delta}$~\cite{hemmati2016emergence}. For example, we can use supercapacitors for short-term power needs and fuel cells for long-term energy needs in a hybrid system. The two parameters~$c$ and~$\bar{\delta}$ may also be set based on a dedicated analysis of the degradation cost, valuation, and long-term planning of the energy storage system. Such modeling is beyond the scope of this work. Still, we use simulations to see the joint effect of the two parameters on the performance of the online algorithms proposed in this paper.

Let $\delta_{t}\geq0$ (in kWh) be the discharge amount at slot~$t$ and~$\bm \delta\in\mathbb{R}^T$ be the discharge vector. \rev{Given the capacity~$c$ and the rate limit~$\bar{\delta}$, we have the capacity constraint~$\sum_{t=1}^{T}\delta_{t}\leq c$ and the discharge constraint~$\delta_{t}\leq\bar{\delta}$, for all~$t\in[T]$.} \rev{For monetary, space, and safety concerns~\cite{Economics_energy_storage}, commercial customers tend not to invest in energy storage systems with huge sizes. Thus, for a large-load user, we can assume that the available capacity never exceeds the total demand (even at its minimum):}
\begin{equation}\label{assumpc}
    c\leq T \underline{d}.
\end{equation}
\rev{For shaving the peak usages of large-load users, we do not consider selling the electricity from the energy storage to the grid, because this practice may increase the demand charge and energy cost instead. In certain cases, one may require using up the available energy storage in the on-peak period to take full advantage of TOU energy arbitrage. In next sections, we first ignore the requirement and discuss, in Section~\ref{Sec_Hybrid}, how to deplete the available energy storage in the period without sacrificing the performance on the peak demand reduction.}

\subsection{Problem Formulation and Optimal Offline Solution} \label{ssec:PDM_formulation}

Now, we formulate the offline peak-minimizing storage-discharging~(PMD) problem:
\begin{equation*}
\begin{split}\label{ProPABD}
  \text{PMD:}~~~\min_{\bm \delta \in \mathbb{R}^T} &~~~~\max_{t\in [T]}\left(d_t-\delta_t\right)\\
  \text{subject to} &~~~~\sum_{t=1}^{T}\delta_t\leq c;~~\text{(Inventory Constraint)}\\
  &~~~~0\leq \delta_t \leq \min\{\bar{\delta},d_t\}\text{, for all }t\in [T].
\end{split}
\end{equation*}
This objective function is the peak usage, i.e., the maximum electricity procurement from the grid. The constraints are due to the storage capacity, the maximum discharging rate, and that there is no need to discharge more than the net demand. This minimization problem has wide application beyond the energy storage context. For example, consider two substitute goods for the same purpose. One is replenished monthly with a fixed order quantity of~$c$ units. The other is replenished daily with flexible order quantities. Our goal is to minimize the highest daily order quantify of the second good~$\max_{t\in[T]}(d_t-\delta_t)$ while fulfilling daily requirements~$d_t,t\in[T]$.

Under the clairvoyant/offline setting where the demand profile~$\bm d$ is known beforehand, the {PMD} problem can be reformulated as a linear program and it is easy to obtain the optimal offline solution as shown in the following proposition.
\begin{proposition}\label{offopt}
    The unique optimal offline solution to the {PMD} problem  is given by

    \centerline{$\delta_t^*=\left[d_t-\max\limits_{t\in [T]}\left[d_t-\bar{\delta}-v\right]^+-v\right]^+\text{, for all }t\in [T],$}
    \noindent where $v \text{ satisfies }\sum_{t=1}^{T}[d_t-v]^+=c.$
\end{proposition}
The above proposition shows that the optimal offline solution can be generated by a constant threshold~--~the maximum of~$\max_{t\in [T]}d_t-\bar{\delta}$ and~$v$ satisfying~$\sum_{t=1}^{T}[d_t-v]^+=c$. Note that Proposition~\ref{offopt} also gives the optimal offline solution to the PRM problem studied in~\cite{mo2021eEnergy}, whose objective function is~$\max_{\bm \delta\in\mathbb{R}^T} \left(\max_{t\in[T]}-\max_{t\in [T]}\left(d_t-\delta_t\right)\right)$ and constraints are the same with PMD.

In practice, however, the net demand profiles of individual customers are usually not given beforehand, as it is challenging to forecast the fluctuating demand and the volatile renewable generation accurately~\cite{jacobsen2012curtailment}. This observation motivates us to consider the more practical online setting where the net demands are revealed sequentially in time, yet one has to make irrevocable discharging decisions at each epoch. This leads to the problem of online peak minimization under energy storage constraints.
The noncumulative nature of the peak objective function and the coupling of online decisions across time slots due to the storage capacity constraint make the problem uniquely challenging. Note that as discussed in Appendix~\ref{app_expdif}, the online PMD problem is fundamentally different from the online PRM problem in~\cite{mo2021eEnergy} with respect to competitive analysis, even though they share the same optimal offline solution characterized in Proposition~\ref{offopt}.
\section{An Optimal Online Algorithm}\label{Sec_OnlineAlg}

In this section, we develop an online algorithm for the PMD problem with the optimal worst-case performance guarantee.




\subsection{Online/Real-Time Setting}

\begin{figure}[t]
  \centering
  \centerline{\includegraphics[width=0.6\columnwidth]{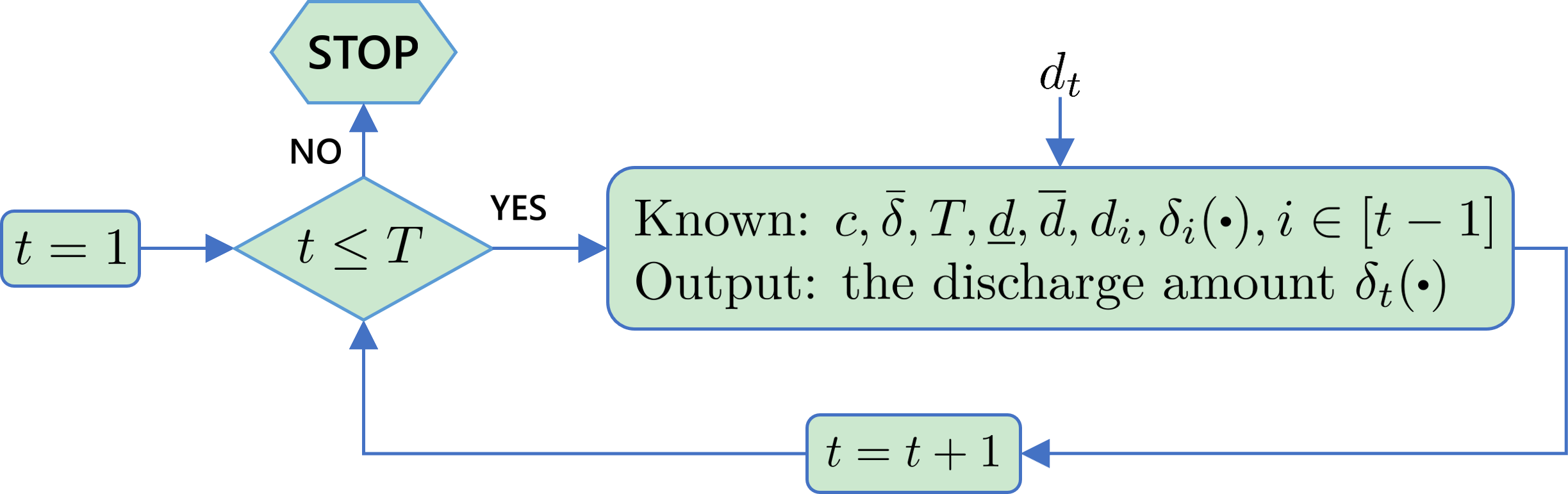}}
  \caption{A flowchart of an online algorithm for PMD. }\label{fig: FlowGraph}
\end{figure}

Under the online/real-time setting, as shown in Fig.~\ref{fig: FlowGraph}, the loads~$\{d_t\}_{t\in[T]}$ are revealed chronologically. At each time $t$, the demand $d_t\in[\underline{d},\overline{d}]$ is revealed and one has to make causal and irrevocable discharging decisions, without knowing future demands. Let~$\mathcal{D}=\{\bm x\in \mathbb{R}^T \mid \underline{d}\leq x_t \leq \overline{d}, \forall t\in[T]\}$.

We employ CR as the performance metric for online algorithms~\cite{borodin2005online}. For a deterministic online algorithm~$\mathfrak{A}$ for {PMD}, CR is defined as the worst-case ratio between the peak usage attained by the algorithm~$\mathfrak{A}$ and that by an optimal offline solution, under all possible input sequences~(or demand profiles), i.e.,
  $CR_{\mathfrak{A}}=\max_{\bm d\in \mathcal{D}}\frac{v_{\mathfrak{A}}(\bm d)}{v(\bm d)}$,
where~$v_{\mathfrak{A}}(\bm d)$ and~$v(\bm d)$ are respectively the objective value of {PMD} under the online algorithm and that under an optimal offline solution with the same demand profile~$\bm d$. \rev{Given a particular input sequence, a deterministic algorithm always produces the same output. In contrast, randomized algorithms throw coins to make decisions during execution. If randomized strategies are allowed where we make decisions at each round following a distribution}, then we can simply extend the definition of CR for a randomized algorithm~$\mathfrak{A}$ by substituting~$v_{\mathfrak{A}}(\bm d)$ with~$\mathbf{E}[v_{\mathfrak{A}}(\bm d)]$, where the expectation is taken with respect to the distributions used by~$\mathfrak{A}$. Clearly, we have~$CR_{\mathfrak{A}}\geq 1$ and prefer online algorithms with smaller CRs. The smallest achievable CR also captures the \emph{price of uncertainty} that quantifies the fundamental performance gap between online and offline decision-makings~\cite{borodin2005online}.

It is known that randomization may help to improve the competitiveness of online algorithms. However, the following proposition, proven in Appendix~\ref{app_rand}, suggests that it suffices to focus on deterministic algorithms for attaining the best CR for the online {PMD}.

\begin{proposition}\label{prop_rand}
  Let~$\mathfrak{A}$ be an arbitrary randomized algorithm and its CR is $CR_{\mathfrak{A}}$. There exists a deterministic online algorithm~$\mathfrak{B}$ whose CR is $CR_{\mathfrak{B}}$ such that~$CR_{\mathfrak{B}}\leq CR_{\mathfrak{A}}$.
\end{proposition}

The above result significantly reduces the design space under consideration for the optimal competitive online algorithm.

\noindent

\subsection{Overview of the \textsf{CR-Pursuit} Framework and Challenges}\label{Sec_CRP} Before proceeding, we review the useful \textsf{CR-Pursuit} algorithmic framework. Usually, one first designs an online algorithm and then computes its CR for evaluating the performance. However, the~\textsf{CR-Pursuit} framework needs reverse thinking. As the name indicates, each algorithm in the framework makes decisions with the goal of ``pursuing'' a specified CR. Specifically, in each decision-making round, the algorithm characterized by a ratio~$\pi$ will choose actions to maintain the online-to-offline objective ratio to be no more than~$\pi$, given the inputs observed so far. Clearly, if we can always choose the actions to maintain the given ratio~$\pi$ under any input sequence, then the ratio~$\pi$ will be the CR of the algorithm. Although the idea behind~\textsf{CR-Pursuit} is simple and intuitive, designing an online algorithm under the framework is non-trivial~\cite{yi2019balancing,lin2019competitive}.


Each algorithm under~\textsf{CR-Pursuit} is characterized by its CR, and we expect to identify the smallest CR which can be maintained under~\textsf{CR-Pursuit}. Here are two key steps of~\textsf{CR-Pursuit}: identifying the best ``pursued'' CR and finding a proper way to maintain the CR in each round. How to carry out these steps is problem-specific and brings unique challenges of applying the~\textsf{CR-Pursuit} framework. More critically, it is not clear whether the best~\textsf{CR-Pursuit} algorithm is also optimal among all online algorithms for the considered problem. If so, \textsf{CR-Pursuit} significantly reduces the search space for the optimal online algorithm to a one-dimensional space over the ratios to maintain. Bearing these challenges in mind, we next devise online algorithms for~{PMD}.

\subsection{First Optimal Online Algorithm for {PMD}}

Recall that~$v(\bm d)$ denotes the optimal offline objective value of {PMD} under the demand~$\bm d$. To apply the~\textsf{CR-Pursuit} framework, we exploit the revealed inputs and identify a reference input sequence~$\bm d^t\in \mathbb{R}^T$ as~$[d_1~d_2~\cdots~d_t~\underline{d}~\cdots~\underline{d}]'$, for each time slot~$t\in[T]$. \rev{These input sequences are based on our most optimistic view of future inputs and selecting them as reference inputs will lead us to optimal online algorithms in terms of CR.} Specifically, we define a class of algorithms, named after~\textsf{pCR-PMD}($\pi$). Each~\textsf{pCR-PMD}($\pi$) is characterized by a ratio~$\pi \geq 1$ specified before observing any inputs. We denote by~$\bm \delta(\pi,\bm d)\in \mathbb{R}^T$ the output sequence under~\textsf{pCR-PMD}($\pi$) and the input sequence~$\bm d$, whose~$t$th element \rev{is independent of future demands~$(d_{t+1},\ldots,d_T)$} and obtained by the formula in Algorithm~\ref{algpCR}. Observe that in each decision-making, we solve an offline PMD under the demand profile~$\bm d^t$, while we pursue the CR~$\pi$ by maintaining the online-to-offline objective ratio under the reference input~$\bm d^t$ to be no more than~$\pi$, for all~$t\in[T]$. However, the question arises whether~$\bm \delta(\pi,\bm d)$ is always a feasible solution to {PMD} for any demand profile~$\bm d\in \mathcal{D}$. \rev{If so, then~\textsf{pCR-PMD}($\pi$) attains the CR~$\pi$ by the definition of CR and the selection of the reference input sequences\footnote{{Note that~$v(\bm d)$ is always lower bounded by~$v(\bm d^t)$, for all~$t\in[T]$. This fact ensures that~$\max_{t\in[T]}(d_t-\delta_t(\pi,\bm d))\leq \pi v(\bm d)$.}}}; otherwise, the CR of~\textsf{pCR-PMD}($\pi$) exceeds~$\pi$. This puzzle motivates us to define and characterize the feasibility of~\textsf{pCR-PMD} algorithms as follows.

\begin{definition}
  \textsf{pCR-PMD}($\pi$) is feasible if~$\bm \delta(\pi,\bm d)$ is a feasible solution to {PMD} for any~$\bm d\in \mathcal{D}$.
\end{definition}

\textsf{pCR-PMD}($\pi$) is feasible if and only if it generates a feasible solution to PMD for any possible input sequence. Then, identifying the best \textsf{pCR-PMD} algorithm is equivalent to finding the smallest ratio~$\pi$ such that \textsf{pCR-PMD}($\pi$) is feasible. Thus, we derive the following proposition, which is proven in Appendix~\ref{app_feasibility} and characterizes the feasibility of~\textsf{pCR-PMD}($\pi$) by a threshold condition.
\begin{proposition}\label{propfeasibility}
  \emph{\textsf{pCR-PMD}($\pi$)} is feasible if and only if~$\Phi(\pi):=\max\limits_{\bm d\in \mathcal{D}}\sum\nolimits_{t=1}^{T}\delta_t(\pi,\bm d)\leq c.$
\end{proposition}
Moreover, the following lemma, proven in Appendix~\ref{app_deinc}, implies that the smallest~$\pi$ with~$\Phi(\pi)\leq c$ gives the best~\textsf{pCR-PMD} algorithm for {PMD}.
\begin{lemma}\label{lemdeinc}
  $\Phi(\pi)$ strictly decreases in~$\pi$ over~$[1,\bar{\pi})$, where~$\bar{\pi}$ is the smallest ratio with~$\Phi(\bar{\pi})=0$.
\end{lemma}

Then, we derive a main result, proven in Appendix~\ref{app_optratio}.  

\begin{algorithm}[t]
\caption{\textsf{pCR-PMD}($\pi$)}\label{algpCR}
\LinesNumbered
  \For{$t=1,2,\ldots,T$} {
  {Identify the reference input sequence~$\bm d^t$ according to observed demands\;
  \textsf{pCR-PMD}($\pi$) requires to discharge the energy storage system by the amount~$\delta_t(\pi,\bm d) = [d_t - \pi  v(\bm d^t)]^+$;}}
\end{algorithm}

\begin{theorem}\label{thmoptratio}
  Given~$c$,~$\bar{\delta}$,~$T$, and~$\mathcal{D}$, the best CR for the online {PMD} is given by the unique solution~$\pi^*$ to the univariate equation~$\Phi(\pi)=c$.
\end{theorem}

The above theorem shows that the best~\textsf{pCR-PMD} algorithm is also the best among all online (deterministic and randomized) algorithms for {PMD}. As a result, we can reduce the search of an optimal online algorithm for {PMD} to the search of a feasible~\textsf{pCR-PMD} algorithm with the smallest possible CR~$\pi^*$. Theorem~\ref{thmoptratio} also characterizes  the price of uncertainty as the unique solution to the univariate equation~$\Phi(\pi)=c$. Nevertheless, the function~$\Phi(\pi)$ seems complicated at first sight and it is unclear how to characterize the optimal CR~$\pi^*$. 


\subsection{Computing the Optimal Competitive Ratio~$\pi^*$}\label{ssec:opt_pi}
\begin{figure*}[t]
  \centering
  \begin{subfigure}[t]{0.32\textwidth}
    \centerline{\includegraphics[width=1\columnwidth]{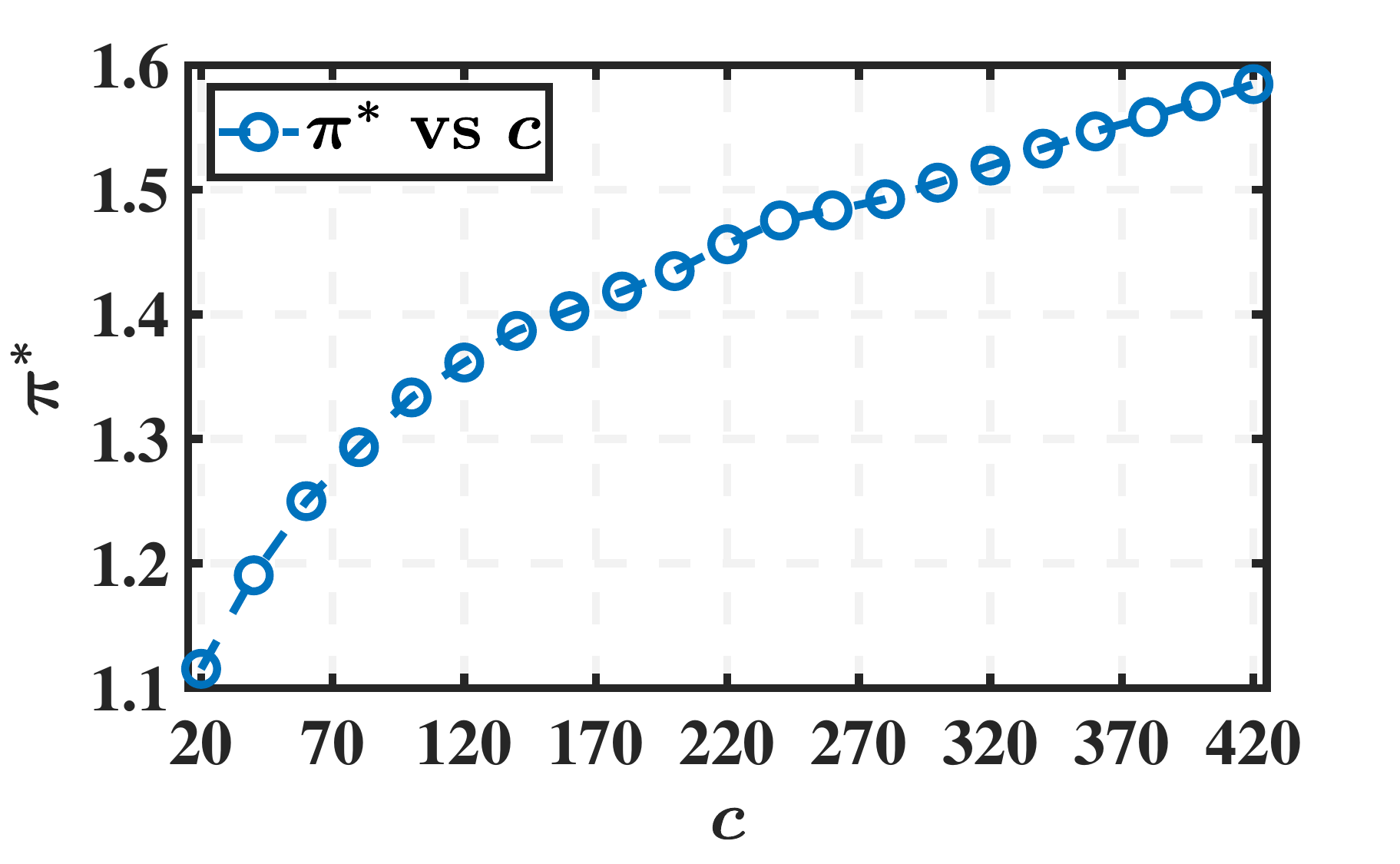}}
  \end{subfigure}
  \begin{subfigure}[t]{0.32\textwidth}
    \leftline{\includegraphics[width=1\columnwidth]{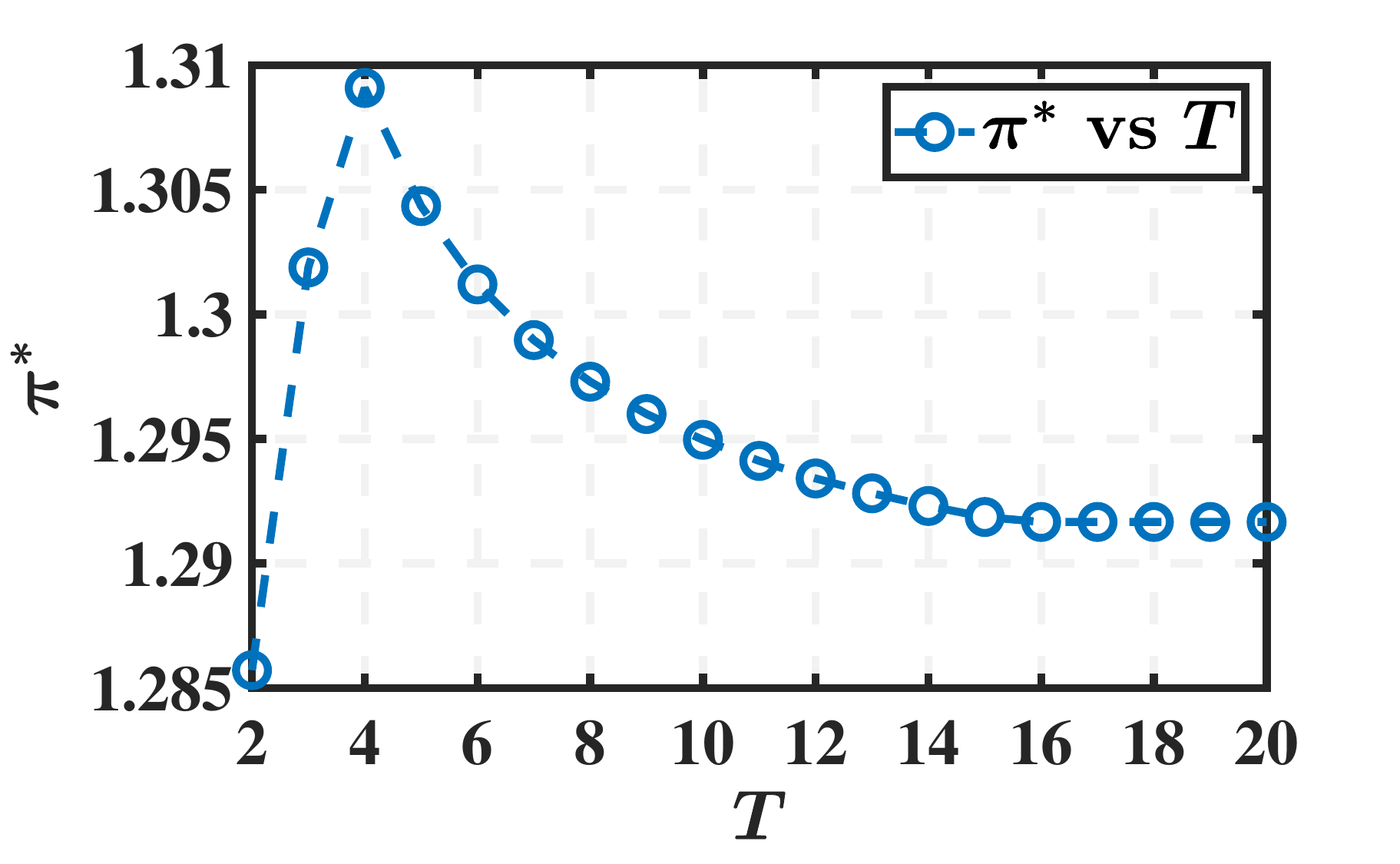}}
  \end{subfigure}
  \begin{subfigure}[t]{0.32\textwidth}
    \centerline{\includegraphics[width=1\columnwidth]{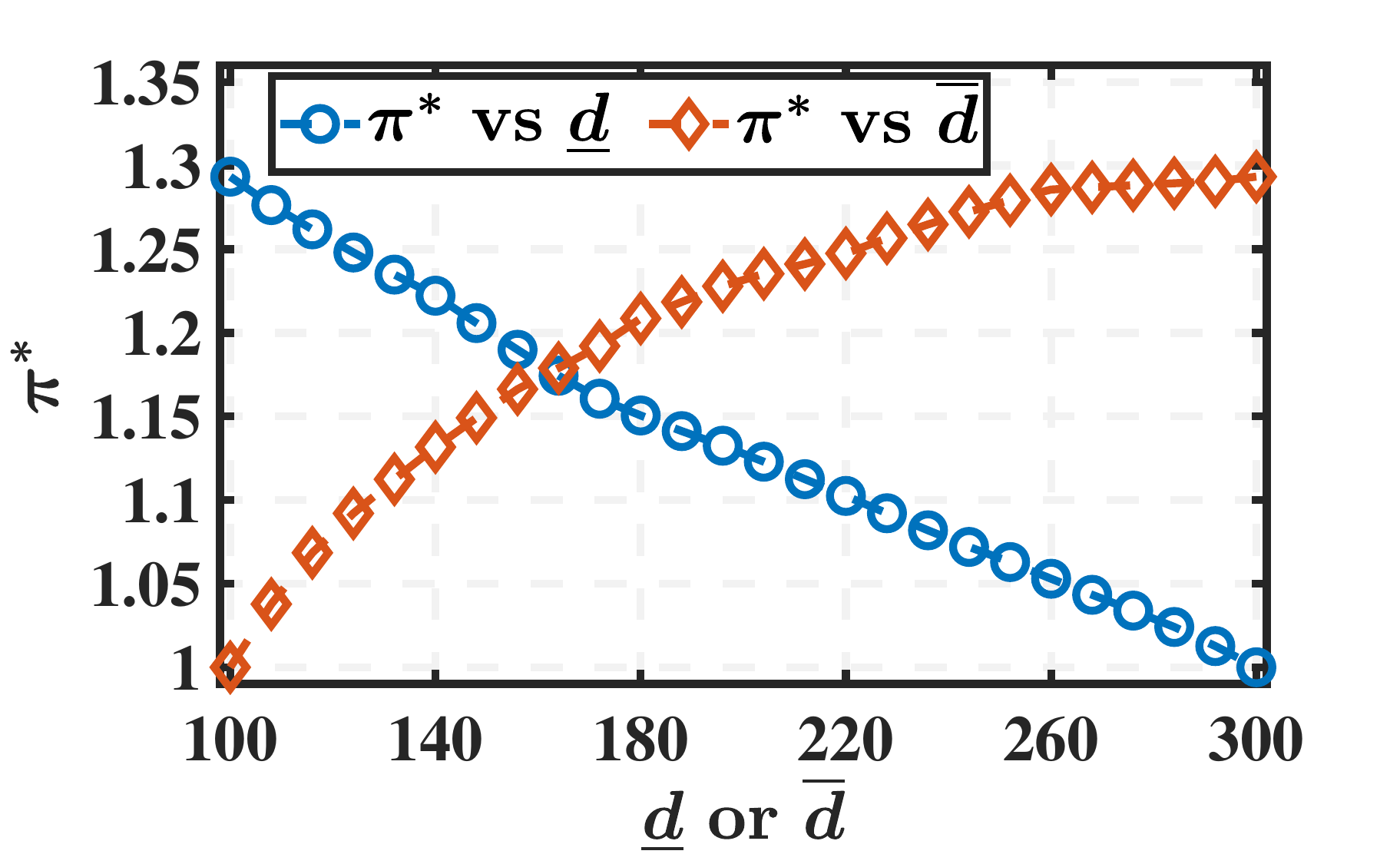}}
  \end{subfigure}
  \caption{The best CR~$\pi^*$ varies as~$c,T,\underline{d}$, and~$\overline{d}$. By default, we set~$c=80$,~$T=12$,~$\underline{d}=100$,~$\overline{d}=300$ and~$\bar{\delta}=\infty$. }
  \label{fig: Sensitivity}
\end{figure*}
Owing to the complexity of~$\Phi(\pi)$, it is almost impossible to derive the analytic expression of the optimal CR~$\pi^*$. Neither is efficient to search for~$\pi^*$ by dynamic programming with a continuous decision space. As a key contribution, we transform the computation of~$\pi^*$ into solving a sequence of linear-fractional problems, each of which is as simple as solving linear programing problems~\cite[Section 4.3.2]{boyd2004convex}. As discussed later, a direct approach leads to an exponential number of linear-fractional problems to solve, regarding the time-slot number~$T$. Interestingly, we carefully exploit the problem structure and show that it suffices to consider at most~$T$ such linear-fractional problems.

First, we observe that if~\textsf{pCR-PMD}($\pi$) is feasible, then for any set~$\mathcal{I}\subseteq [T]$ and any input sequence~$\bm d\in \mathcal{D}$, we have~$\sum_{i\in \mathcal{I}}\!\left(d_i-\pi  v(\bm d^i)\right)\!\leq\! \sum_{t=1}^{T}\![d_t - \pi v(\bm d^t)]^+\leq c$, implying
\begin{equation}\label{lfinequality}
  \frac{\sum_{i\in\mathcal{I}}d_i-c}{\sum_{i\in \mathcal{I}}v(\bm d^i)}\leq \pi.
\end{equation}
That is, the best CR $\pi^*$ is lower bounded by the smallest ratio~$\pi$ satisfying~(\ref{lfinequality}) for any set~$\mathcal{I}\subseteq [T]$ and any profile~$\bm d\in \mathcal{D}$. Moreover, there exist an index set~$\mathcal{I}^*$ and a demand profile~$\bm d^*$ such that the equality of~(\ref{lfinequality}) holds for~$\mathcal{I}^*$,~$\bm d^*$ and~$\pi^*$; thus the lower bound is also tight. This observation, together with  Lemma~\ref{lemdeinc} and Proposition~\ref{propfeasibility}, leads to the following proposition.
\begin{proposition}\label{proppaiexp}
Given~$c$,~$\bar{\delta}$,~$T$, and~$\mathcal{D}$, the best CR for the online {PMD} is given by~$$\max_{\mathcal{I}\subseteq [T],\bm d\in \mathcal{D}}\frac{\sum_{i\in\mathcal{I}}d_i-c}{\sum_{i\in \mathcal{I}}v(\bm d^i)}.
$$
\end{proposition}

The above proposition suggests that we can find the optimal CR~$\pi^*$ by solving a sequence of linear-fractional programs \rev{before implementing \textsf{pCR-PMD}($\pi^*$).} Specifically, we define a class of linear-fractional problems. \rev{Each problem is parameterized by a nonempty set~$\mathcal{I}$ that is a subset of the set~$[T]$} and named after~\textsf{CR-Compute}($\mathcal{I}$):
\begin{equation*}
\begin{split}
  \textsf{CR-Compute}(\mathcal{I}):~~~&~~~~\max_{\rev{\bm x}\in \mathcal{D},u_i,\delta_{ij}}~~\frac{\sum_{i\in\mathcal{I}}\rev{x_i}-c}{\sum_{i\in \mathcal{I}}u_i}\\
      \text{subject to}~&~~~~\sum_{j=1}^{T}\delta_{ij}=c\text{, for all }i\in[T];~~0\leq \delta_{ij}\leq \bar{\delta}\text{, for all }i,j\in [T];\\
      &~~~~\rev{x_j}-\delta_{ij}\leq u_i\text{, for all }1\leq j\leq  i\leq T;\\
      &~~~~\underline{d}-\delta_{ij}\leq u_i\text{, for all }1\leq i< j\leq T.
    \end{split}
  \end{equation*}
The constraints above are due to the offline {PMD} problem solved in each time slot under any~\textsf{pCR-PMD} algorithm. Specifically, the variables~$\delta_{ij}, j\in[T]$ and the variable~$u_i$ are respectively associated with the optimal solution and the optimal objective of {PMD} under \rev{the reference input sequence~$[x_1~x_2~\cdots~x_i~\underline{d}~\cdots~\underline{d}]'$}, for all~$i\in[T]$. The objective function comes from Proposition~\ref{proppaiexp}. By abuse of notation, let \textsf{CR-Compute}($\mathcal{I}$) denote the resulting optimal objective value. \rev{We note that \textsf{CR-Compute}($\mathcal{I}$) depends solely on the prior information~$(c,T,\underline{d},\overline{d})$ and the set~$\mathcal{I}$.} Moreover, by above analysis, we easily attain the following lemma.
\begin{lemma}\label{lemtolf}
  For any nonempty set~$\mathcal{I}\subseteq [T]$, it follows that~$\textsf{CR-Compute}(\mathcal{I})=\max_{\bm d\in \mathcal{D}}\frac{\sum_{i\in\mathcal{I}}d_i-c}{\sum_{i\in \mathcal{I}}v(\bm d^i)}.$
\end{lemma}

We conclude from Proposition~\ref{proppaiexp} and Lemma~\ref{lemtolf} that the best CR~$\pi^*$ of all online algorithms for {PMD} is given by~$\max_{\mathcal{I}\subset [T]}\textsf{CR-Compute}(\mathcal{I})$. That is, we can convert the computation of~$\pi^*$ into solving a sequence of linear-fractional programs~\textsf{CR-Compute}($\mathcal{I}$) parameterized by~$\mathcal{I}\in [T]$. Note that, for any~$\mathcal{I}\subset [T]$,~\textsf{CR-Compute}($\mathcal{I}$) can be tackled by transforming to a linear program~\cite[Section 4.3.2]{boyd2004convex}. Although Proposition~\ref{proppaiexp} and Lemma~\ref{lemtolf} suggest a numerical way to obtain the best CR~$\pi^*$, we have to solve an exponential number~($2^T$) of linear programs for the best CR~$\pi^*$. That is undesirable and we expect to exclude as many redundant~\textsf{CR-Compute}($\mathcal{I}$) programs as possible. To this end, we exploit the worst-case input sequences of~\textsf{pCR-PMD}($\pi^*$), which will use up the storage under the algorithm.


\begin{lemma}\label{lemparwc}
  There exist an index~$\tau\in[T]$ and~$\bm d\in \mathcal{D}$ with~$\sum_{t=1}^{T}\delta_t(\pi^*,\bm d)=c$ such that~\mbox{$\delta_t(\pi^*,\bm d)>0$} if~$t\in [\tau]$ and~$\delta_t(\pi^*,\bm d)=0$ otherwise.
\end{lemma}

The above lemma, proven in Appendix~\ref{app_parwc}, indicates the existence of a particular worst-case input sequence such that~\textsf{pCR-PMD}($\pi^*$) will always discharge certain energy until it uses up all the stored energy. Combining Lemma~\ref{lemtolf}, Lemma~\ref{lemparwc}, and Proposition~\ref{proppaiexp}, we attain the following theorem. It states that we can significantly reduce the number of~\textsf{CR-Compute}($\mathcal{I}$) programs for computing the best CR to a linear number (less than~$T$) instead of an exponential number~($2^T$) in terms of the time-slot number~$T$. 
\begin{theorem}\label{thmoptratiocal}
Given~$c$,~$\bar{\delta}$,~$T$, and~$\mathcal{D}$, the best CR for the online {PMD} is given by

 \centerline{$\pi^*=\max\limits_{\mathcal{I}\in\{[t]\mid t=\tau+1,\ldots,T\}}~ \textsf{CR-Compute}(\mathcal{I})\text{, where }\tau=\lfloor c/\overline{d}\rfloor.
$}
\end{theorem}

\rev{So far, we have shown how to obtain the best possible CR~$\pi^*$ by solving a linear number of linear-fractional programs before the operation period, solely based on the prior information~$(c,T,\underline{d},\overline{d})$.} In Fig.~\ref{fig: Sensitivity}, we illustrate~$\pi^*$ over~$(c,T,\underline{d},\overline{d})$ with a large enough~$\bar{\delta}$ such that the rate constraints will not be active. The impact of~$\bar{\delta}$ will be discussed later in simulations based on real-world traces. As we can see, the optimal CR~$\pi^*$ decreases as the storage capacity or the demand uncertainty decreases (i.e., the set of possible demand profiles shrinks), but the change of~$\pi^*$ is not monotone as~$T$ varies. Intuitively, a larger~$T$ leads to a higher dimension of the input sequence space, making the online decision-making harder. This may account for the initial increase of~$\pi^*$ in~$T$. On the other hand, as~$T$ increases, the optimal offline outcomes become worse and worse. Then, the possible improvement of the relative performance between the online and the optimal offline outcomes may explain why the optimal CR~$\pi^*$ decreases in~$T$ after~$T$ exceeds a certain value.  

\section{Anytime-Optimal~\textsf{pCR-PMD} Algorithm}\label{Sec_Adaptive}
\textsf{pCR-PMD}($\pi^*$) maintains the same performance ratio~$\pi^*$ for any possible input sequence. We expect an online algorithm with the optimal worst-case performance and an adaptive average-case performance. To this end, we extend the~\textsf{pCR-PMD}($\pi^*$) algorithm to an anytime-optimal one by adaptively pruning the input space based on the observations so far and adjusting the online
decisions for the best competitiveness regarding the residual uncertainty. That is, the anytime-optimal algorithm captures the varying prices of future uncertainty and pursues the adaptive best possible CR at any epoch to improve online decisions, given the inputs and online actions so far. In this way, we retain the optimal worst-case performance and achieve an adaptive worst-case performance. These results suggest the potential of combining efficiency and robustness, leading to more practical competitive online algorithms. Moreover, we will see that the adaptive approach in this work is more insightful, robust, and efficient in principle than the counterpart in~\cite{mo2021infocom} by considering the anytime-optimal competitiveness.  

Before proceeding, we extend the concept of CR to a sequence of anytime CRs. Recall that the CR of an algorithm implies the worst-case online-to-offline ratio of the objective value attained by the algorithm over all possible inputs. Intuitively, we say that the anytime CR at slot~$t$ of an algorithm for the online PMD indicates the worst-case online-to-offline ratio of peak usage that is attained by the algorithm, after observing the specific inputs until slot~$t$.  Specifically, given revealed inputs~$d_k,k\in[t]$, the anytime CR at time slot~$t$ of an online algorithm~$\mathfrak{A}$ is defined as~$\pi^\mathfrak{A}_t = \max_{\bm x\in \mathcal{D}_t}\frac{v_{\mathfrak{A}}(\bm x)}{v(\bm x)}$, where~$\mathcal{D}_t=\{\bm x\in\mathcal{D}\mid x_k=d_k, \forall k\in[t]\}$. In the following, we further introduce the concept of anytime-optimal CR to characterize the best competitive online algorithm for PMD, given the observed inputs and online actions so far.
\begin{definition}
  Given inputs~$d_k,k\in[t]$ and online actions~$\delta_k,k\in[t\!-\!1]$, the anytime-optimal CR at time slot~\mbox{$t\in[T]$} is defined as
  $\pi^*_t = \min_{\mathfrak{A}\in \mathcal{A}_t}~\max_{\bm x\in \mathcal{D}_t}\frac{v_{\mathfrak{A}}(\bm x)}{v(\bm x)}$,
where the set~$\mathcal{A}_t$ consists of online algorithms~$\mathfrak{A}$ satisfying~\mbox{$\delta_k(\mathfrak{A},\bm x)=\delta_k$}, for all~$k\in[t-1]$ and~$\bm x\in \mathcal{D}_{t-1}$.
\end{definition}

\begin{algorithm}[t]
\caption{Anytime-Optimal \textsf{pCR-PMD} Algorithm}\label{algadaptivepCR}
  \For{$t=1,2,\ldots,T$}{
  {Update the ratio~$\pi^*_t$ according to Algorithm~\ref{algadaptivepCRcompute}\;}
  {The discharge amount at time slot~$t$ is given by~$\delta_t = [d_t - \pi^*_t   v(\bm d^t)]^+$\;
 }
 }
\end{algorithm}
  \begin{algorithm}[t]
\caption{A Bisection Method for~$\pi^*_t$ in the Anytime-Optimal \textsf{pCR-PMD} Algorithm}\label{algadaptivepCRcompute}
  \KwIn{$c,\bar{\delta},T,\underline{d},\overline{d}$, observed inputs~$d_k,k\in[t]$, implemented actions~$\delta_k,k\in[t-1]$, and~$\pi^*_{t-1}$;}
  \KwOut{The anytime-optimal CR at time slot~$t$ under Anytime-Optimal \textsf{pCR-PMD}: $\pi^*_t$;}
  {$\pi_{lb}=\max_{k\in[t-1]}(d_k-\delta_k)/v(\bm d^t)$, $\pi_{ub}=\pi^*_{t-1}$, $q = \max_{\mathcal{I}\in \mathcal{I}_t} \textsf{AOCR-THR}(\pi_{lb},\mathcal{I})$;}\\
  \If{$q\leq (c-\sum_{k=1}^{t-1}\delta_k)$} {$\pi^*_t=\pi_{lb}$, \Return;}
  \While{$\pi_{ub}-\pi_{lb}\geq \epsilon$}{
  {$\pi=(\pi_{lb}+\pi_{ub})/2$, $q = \max_{\mathcal{I}\in \mathcal{I}_t} \textsf{AOCR-THR}(\pi,\mathcal{I})$\;}
   {
  {$\begin{cases}
\pi_{lb}=\pi & \text{if } q > (c-\sum_{k=1}^{t-1}\delta_k);\\
    \pi_{ub}=\pi &\text{otherwise};
\end{cases}$}
}
  }
\end{algorithm}

While the best possible CR~$\pi^*$ captures the price of uncertainty before we know any inputs, the anytime-optimal CR~$\pi^*_t$ at slot~$t$ characterizes the instantaneous price of uncertainty at time slot~$t$, given the observed inputs~$d_k,k\in[t]$ and implemented actions~$\delta_k,k\in[t-1]$. That is, an online algorithm can at best maintain the online-to-offline ratio of peak purchased demand to be~$\pi^*_t$, for all possible future inputs. It is clear that~$\pi^*_t$ is subject to the observed inputs and online actions, for all~$t\in[T]$. Based on the introduction of anytime-optimal CRs, we next introduce the adaptive extension of the~\textsf{pCR-PMD}($\pi^*$) algorithm.

The anytime-optimal~\textsf{pCR-PMD} algorithm improves~\textsf{pCR-PMD}($\pi^*$) following the anytime-optimal competitive framework, where we always pursue the best possible worst-case performance guarantee at any time slot based on previous inputs and online decisions. That is, at each time slot~$t$, instead of the optimal CR~$\pi^*$, we maintain the online-to-offline ratio of the peak purchased demand in future time slots to be no more than the anytime-optimal CR~$\pi^*_t$, since~$\pi^*_t$ essentially captures the price of future uncertainty given the revealed inputs and implemented actions so far. We present the pseudocodes of the anytime-optimal~\textsf{pCR-PMD} in Algorithm~\ref{algadaptivepCR}. Note that it is unnecessary to decrease the purchased demand of the current time slot to be less than the peak purchased demand in previous time slots. This fact brings additional difficulty in characterizing the anytime-optimal CRs, compared to the best CR~$\pi^*$, as elaborated in the sequel.

\rev{First, we observe that the anytime-optimal CR~$\pi^*_1$ is no more than the best CR~$\pi^*$~($\pi^*_1\leq \pi^*$), because the former absorbs~$d_1$ and involves less uncertainty. Similarly, we can obtain~$\pi^*_1$ by solving \textsf{CR-Compute} programs with an additional constraint~$x_1=d_1$.} Second, since the anytime-optimal~\textsf{pCR-PMD} algorithm makes decisions by pursuing the anytime-optimal CR at each time slot, we conclude that the sequence~$\pi^*_t,t\in[T]$ is nonincreasing in~$t$, whatever the inputs are. Third, we see that~$\pi^*_t\geq \max_{k\in[t-1]}(d_k-\delta_k)/v(\bm d^t)$, because the ultimate online peak usage under the input sequence~$\bm d^t$ is no less than~\mbox{$\max_{k\in[t-1]}(d_k-\delta_k)$}. If the equality holds, we have~$\delta_t=[d_t-\max_{k\in[t-1]}(d_k-\delta_k)]^+$. Based on the three observations, we shall show how to search for~$\pi^*_t$ by a bisection method. To this end, in the following, given observed inputs~$d_k,k\in[t]$ and actions~$\delta_k,k\in [t-1]$, we define a linear program parameterized by a ratio~$\pi\in [\max_{k\in[t-1]}(d_k-\delta_k)/v(\bm d^t),\pi^*_{t-1}]$ and a set~$\mathcal{I}\in \mathcal{I}_t$, where~$\mathcal{I}_t=\{[k]\backslash [t]\mid k=t,t+1,\ldots,T\}$, denoted as
 $\textsf{AOCR-THR}(\pi,\mathcal{I})$, as follows:
\begin{equation*}
    \begin{split}
      \max_{u_i,\delta_{ij}, \rev{x_i},i\in\mathcal{I}}~ &\hspace{0em}\left[d_t-\max\left\{\pi v(\bm d^t),\max_{k\in[t-1]}(d_k-\delta_k)\right\}\right]^++\sum_{i\in\mathcal{I}}(\rev{x_i}-\pi u_i)\\
      \text{subject to}~&~~~\rev{x_i=d_i\text{, for all }i\in[t];}\\
      &~~~\sum_{j=1}^{T}\delta_{ij}=c\text{, for all }i\in \mathcal{I};~~0\leq \delta_{ij}\leq \bar{\delta}\text{, for all }i\in \mathcal{I}\text{ and }j\in [T];\\
      &~~~\rev{x_j}-\delta_{ij}\leq u_i\text{, for all }i\in \mathcal{I}\text{ and } j\leq i;~~\underline{d}-\delta_{ij}\leq u_i\text{, for all }t\leq i< j\leq T;\\
      &~~~\max_{k\in[t-1]}(d_k-\delta_k)\leq \rev{x_i},\;\underline{d}\leq \rev{x_i}\leq \overline{d},~~\max_{k\in[t-1]}(d_k-\delta_k)\leq \pi u_i\text{, for all }i\in \mathcal{I}.
    \end{split}
  \end{equation*}

The above objective function corresponds to the sum of discharge amounts over a set of time slots assuming that we uniformly maintain the online-to-offline ratio of peak purchased demand to be~$\pi$ in all future time slots. Similar to~\textsf{CR-Compute}($\mathcal{I}$), the constraints of~$\textsf{AOCR-THR}(\pi,\mathcal{I})$ are due to the offline {PMD} problem solved in each time slot under the anytime-optimal~\textsf{pCR-PMD}, while the difference lies in that we never decrease the purchased demand of a future time slot to be less than the observed peak purchased demand, as described by the inequality constraints on the last line. By similar arguments for computing the best CR~$\pi^*$, we set~$\pi^*=\pi^*_0$ and conclude that the anytime-optimal CR at time slot~$t$ should be the smallest ratio~$\pi$ in~$[\max_{k\in[t-1]}(d_k-\delta_k)/v(\bm d^t),\pi^*_{t-1}]$ such that~$\max_{\mathcal{I}\in \mathcal{I}_t} \textsf{AOCR-THR}(\pi,\mathcal{I})$ does not exceed a threshold, namely, the inventory~$c-\sum_{k=1}^{t-1}\delta_k$. Thus, we can search for~$\pi^*_t$ by the bisection method as shown in Algorithm~\ref{algadaptivepCRcompute}. Note that the anytime-optimal CRs in this work and~\cite{mo2021eEnergy} are different, as exemplified in Appendix~\ref{app_expdif}.

If the input sequence~$\bm d$ is in the worst case regarding~\textsf{pCR-PMD}($\pi^*$), then~\mbox{$\pi^*_t=\pi^*$}, for all~\mbox{$t\in[T]$}. Otherwise, there is an index~$\tau\in[T]$ such that~$\pi^*_t< \pi^*$, for all~$t\geq \tau$. From this perspective, we show that the anytime-optimal~\textsf{pCR-PMD} still achieves the optimal CR among all online algorithms for~{PMD}, and performs better than~\textsf{pCR-PMD}($\pi^*$) under general cases. As a whole, extending~\textsf{pCR-PMD}($\pi^*$) to the anytime-optimal~\textsf{pCR-PMD} can improve the practical utilization of storage for minimizing the peak demand, which will be further verified by real-world traces later. 

\rev{
\section{Discussions and Extensions}\label{Sec_Hybrid}
\subsection{Monthly Demand Charge}\label{DaytoMonth}
So far, we have obtained optimal online algorithms for reducing the peak usage in the on-peak period of a day. While this practice suits the daily demand charge~\cite{Daily_DM,vaghefi2014hybrid} and is of independent interest for peak shaving, a typical billing cycle for demand charges is a calendar month. This fact means that we should take the peak usage over the previous days in the month into account when dealing with the on-peak period of a certain day. We achieve this purpose via the similar idea of deriving the anytime-optimal~\textsf{pCR-PMD} algorithm. Therefore, we will not discharge the energy storage such that the purchased demand is less than the current peak usage. Specifically, let~$d^{op}$ be the current peak usage of the month and we adapt $\textsf{AOCR-THR}(\pi,\mathcal{I})$ by adding a constraint~$d^{op}\leq \pi u_i\text{, for all }i\in \mathcal{I}.$ As in~Algorithm~\ref{algadaptivepCRcompute}, we apply a bisection procedure to obtain the anytime-optimal ratio~$\pi^*_t$ with slightly changing the initialization as~$\pi_{lb}=\max\{d^{op},\max_{k\in[t-1]}(d_k-\delta_k)\}/v(\bm d^t).$ After that, we implement Algorithm~\ref{algadaptivepCR} as before.}


\rev{
\subsection{Depleting Cheaper Energy Storage}
We observe that the anytime-optimal~\textsf{pCR-PMD} algorithm may not use up the available storage at the end of the operating on-peak period. However, the energy storage is often cheaper than the electricity purchased in on-peak periods. For example, we may store the energy storage in batteries which we charge in off-peak periods with a lower electricity rate. Motivated by this observation, we can adapt the anytime-optimal~\textsf{pCR-PMD} by discharging the maximum possible amount~$\overline{\delta}_t$ of stored energy for maintaining the online-to-offline ratio to be no more than CR~$\pi^*_t$, instead of the minimum amount~$\underline{\delta}_t$ for each slot~\mbox{$t\in[T]$}. In this way, we can fully consume the available energy storage in the operating period for every possible input sequence while maintaining the same anytime-optimal CRs.

Precisely, we set the discharge amount~$\delta_t=\overline{\delta}_t=\underline{\delta}_t+\left(c-\sum_{k=1}^{t-1}\delta_k-\max_{\mathcal{I}\in \mathcal{I}_t} \textsf{AOCR-THR}(\pi^*_t,\mathcal{I}) \right),$
where~$\underline{\delta}_t=[d_t - \pi^*_t  v(\bm d^t)]^+$ is the discharge amount in the original anytime-optimal~\textsf{pCR-PMD} algorithm. The additional discharge amount~$\overline{\delta}_t-\underline{\delta}_t$ is equal to the redundant storage capacity that will not be useful in maintaining the anytime-optimal CR~$\pi^*_t$, regardless of the future demands. Meanwhile, we identify the occurrence of the redundancy when~$\pi^*_t$ equals the initial value of~$\pi^*_{lb}$ in Algorithm~\ref{algadaptivepCRcompute}. In this case, the anytime-optimal CR~$\pi^*_t$ is subject to the peak usage caused by the conservative decisions in previous time slots.}

\rev{
\subsection{Cost-Minimization over An On-Peak Period}
Apart from the peak usage, the user also concerns the energy consumption over the operation period, especially under a tariff with daily demand charges and RTP pricing~\cite{Daily_DM,vaghefi2014hybrid}. Thus, to reduce the electricity cost, we formulate the cost-minimizing storage-discharging (CMD) \mbox{problem in the period as}
\begin{equation*}
\begin{split}\label{ProhPABD}
  \text{CMD:}~~~~~\min_{\bm \delta \in \mathbb{R}^T} &~~~~\sum_{t=1}^{T}w^e_t\left(d_t-\delta_t\right)+w^p\max_{t\in [T]}\left(d_t-\delta_t\right)\\
  \text{subject to} &~~~~\sum_{t=1}^{T}\delta_t\leq c;~~0\leq \delta_t \leq \min\{\bar{\delta},d_t\}\text{, for all }t\in [T],
\end{split}
\end{equation*}
where~$w^e_t$ and~$w^p$ are the weighting factors corresponding to the energy consumption (in kWh) of slot~$t$ and the peak usage (in kWh) over the~$T$ slots, respectively. A simple choice of the factors is to associate them with the price rates. That is, let~$w^e_t$ and~$w^p$ respectively be the volume charge rate of slot~$t$ and the daily demand charge rate. In this case, we usually attain the following assumption\footnote{\rev{A daily demand-charge rate~(around~$10$~\$/kW) is typically significantly higher than the RTP rate differences (around~$0.2$~\$/kWh)~\cite{Daily_DM}, while the daily on-peak period has at most~$T=96$ slots (namely, $24$~hours) when the measurement interval is 15 minutes.}}:
\begin{equation}\label{assumpp}
    w^p\geq T \max_{i,j\in[T]}\left(w^e_i-w^e_j\right).
\end{equation}

We note that the problem CMD differs from the problem PMD formulated in Sec.~\ref{ssec:PDM_formulation} only in the objective function. Under Assumption~(\ref{assumpp}), we observe that CMD and PMD share the same optimal (offline) solution, as given in Proposition~\ref{offopt}. Nevertheless, there are respective challenges to use competitive analysis to design online algorithms for the two problems. Let us assume that Assumption~(\ref{assumpp}) holds hereinafter.}

Given the similarity between CMD and PMD, one way to solve the CMD problem in an online fashion is to leverage the schemes developed for PMD. Specifically, given the same demands revealed sequentially in time, one can apply the~\textsf{pCR-PMD}($\pi^*$) and  anytime-optimal~\textsf{pCR-PMD} algorithms to obtain an online solution for problem PMD, for minimizing the peak-demand charge, and then use the same solution for solving problem CMD, for minimizing the weighted electricity cost. Interestingly, the following lemma shows that doing so gives us the first set of online algorithms for minimizing the electricity cost with a strong performance guarantee.
\begin{lemma}\label{lemhPAD}
  By directly applying the solutions from \textsf{pCR-PMD}($\pi^*$) (for solving problem PMD) to problem CMD, we achieve a CR no larger than $\pi^*$. Similarly, the achieved anytime CR of slot~$t$ under the anytime-optimal~\textsf{pCR-PMD} (for solving problem CMD) is no larger than~$\pi^*_t$, for all~$t\in[T]$.
\end{lemma}
The proof comes from the definition of (anytime-optimal) CRs and the fact that problems CMD and PMD share the same optimal offline solution given the same~$\bm d$ under the assumption in~\eqref{assumpp}. The lemma says that the two peak-minimizing online algorithms can be used to generate online solutions for CMD, with a strong worst-case performance guarantee. However, similar results cannot be obtained if we consider online peak-reduction and cost-reduction maximization problems under the metric of competitive ratio.

\rev{Similarly, we can tactfully use up the available energy storage for the horizon and further reduce the energy cost. Meanwhile, it is of great interest to design algorithms with better CRs for CMD and consider reducing the electricity cost over a month, which we leave for future studies.}

\rev{\subsection{Charging/Discharging Setting}
For battery energy storage systems, one may expect to recharge the batteries in the operation horizon, though this practice may increase the degradation cost~\cite{he2015optimal,xu2016modeling,foggo2017improved}. In the following, we shall see that the peak-minimization problem under the charging/discharging setting is mathematically different from the one considered here, solving which requires us to address a different set of theoretical challenges and thus is beyond the scope of this paper.

First, the consumer-owned renewable generation can charge the energy storage system. We should remodel the demand uncertainty and determine the usage of the renewable generation at each slot. This may lead to online peak minimization under dynamic inventory constraints and unknown replenishment. Second, the recharging operation brings additional difficulty in shaving peaks with energy storage. For example, we cannot simply scale the available capacity by the discharge efficiency ratio before the problem formulation as we have done in this work, because we should consider the charge and discharge efficiencies simultaneously. Third, it is harder to make online decisions when we also consider the energy cost, because it may happen that recharging does not help reduce the peak usage but increases the energy cost. Overall, it remains a  largely open problem to design online algorithms with decent performance guarantees under the charging/discharging setting and we leave it to the future work.

Nevertheless, it is worth mentioning that our results for the discharging-only setting can serve as a nontrivial benchmark for future studies of the charging/discharging setting.

}

\section{Simulation}\label{Sec_Simulation}

\subsection{Data and Evaluation Setup}\label{ssec_sim_setup}

\begin{figure}[t]
  \centering{\includegraphics[width=.7\textwidth]{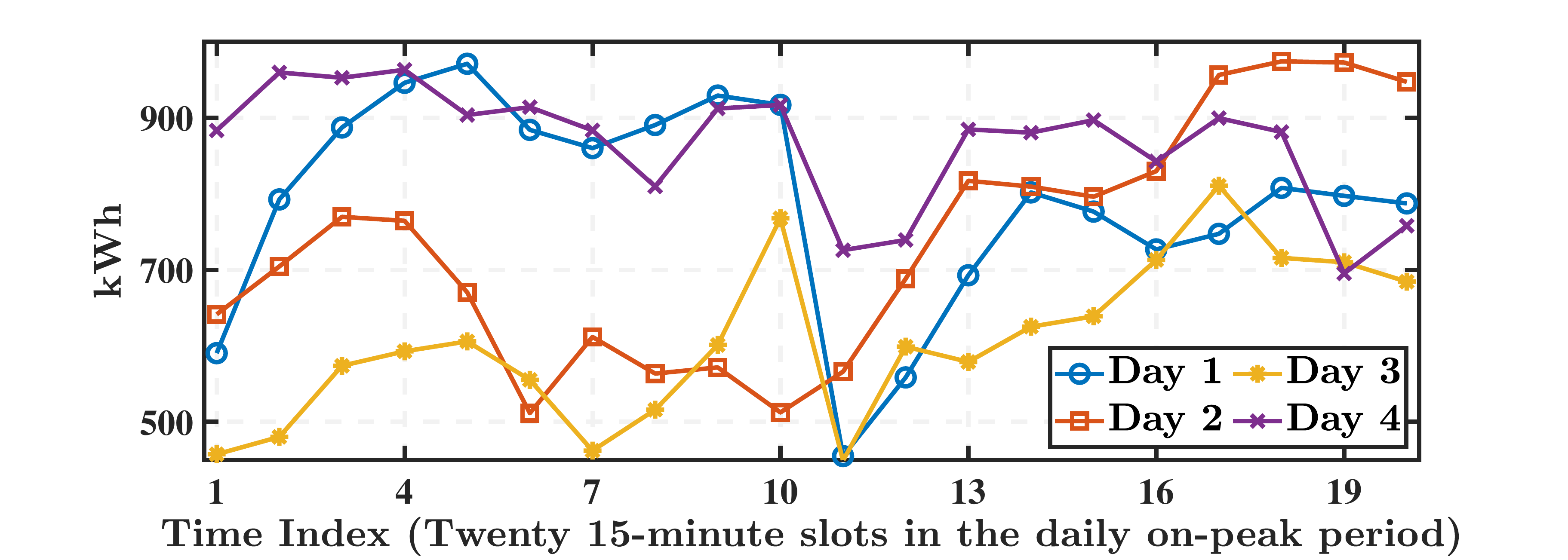}}
  \caption{An illustration of load volatility via four representative demand curves from real-world traces.}\label{fig: Demand}
\end{figure}

We conduct simulations by using $3$-month traces from an EV charging station. The raw data consist of a collection of transactions. Each transaction records the starting time, duration (in minutes), and consumed energy (in~kWh) of an ordered charging. Assume that the charging rate is constant in the recorded duration and convert the raw data into time series regarding the electricity consumption of the station. Then, we can identify the on-peak period in a day and uniformly divide the period into~$15$-minute time slots. Note that the demand from EV charging relies on the number of available charging piles and is scalable. Thus, we slightly scale up the demand such that the expected power exceeds~$3$~MW~($750$ kWh for a~$15$-minute duration), which refers to large power in many tariffs~\cite{Daily_DM}. Then, on a daily basis, we preprocess the raw data and evaluate the performance of online algorithms. Specifically, we divide the daily on-peak period into~$T=20$ time slots and set the demand bounds as~$(\underline{d},\overline{d})=(442.91,1020.10)$~kWh. We plot four demand curves in Fig.~\ref{fig: Demand} and observe that the EV charging loads are highly volatile.

\subsection{Empirical Evaluation of Online Algorithms}

\subsubsection{Anytime-Optimal~\textsf{pCR-PMD} vs.~\textsf{pCR-PMD}($\pi^*$)}

\begin{figure}[t]
    \centerline{\includegraphics[width=0.6\columnwidth]{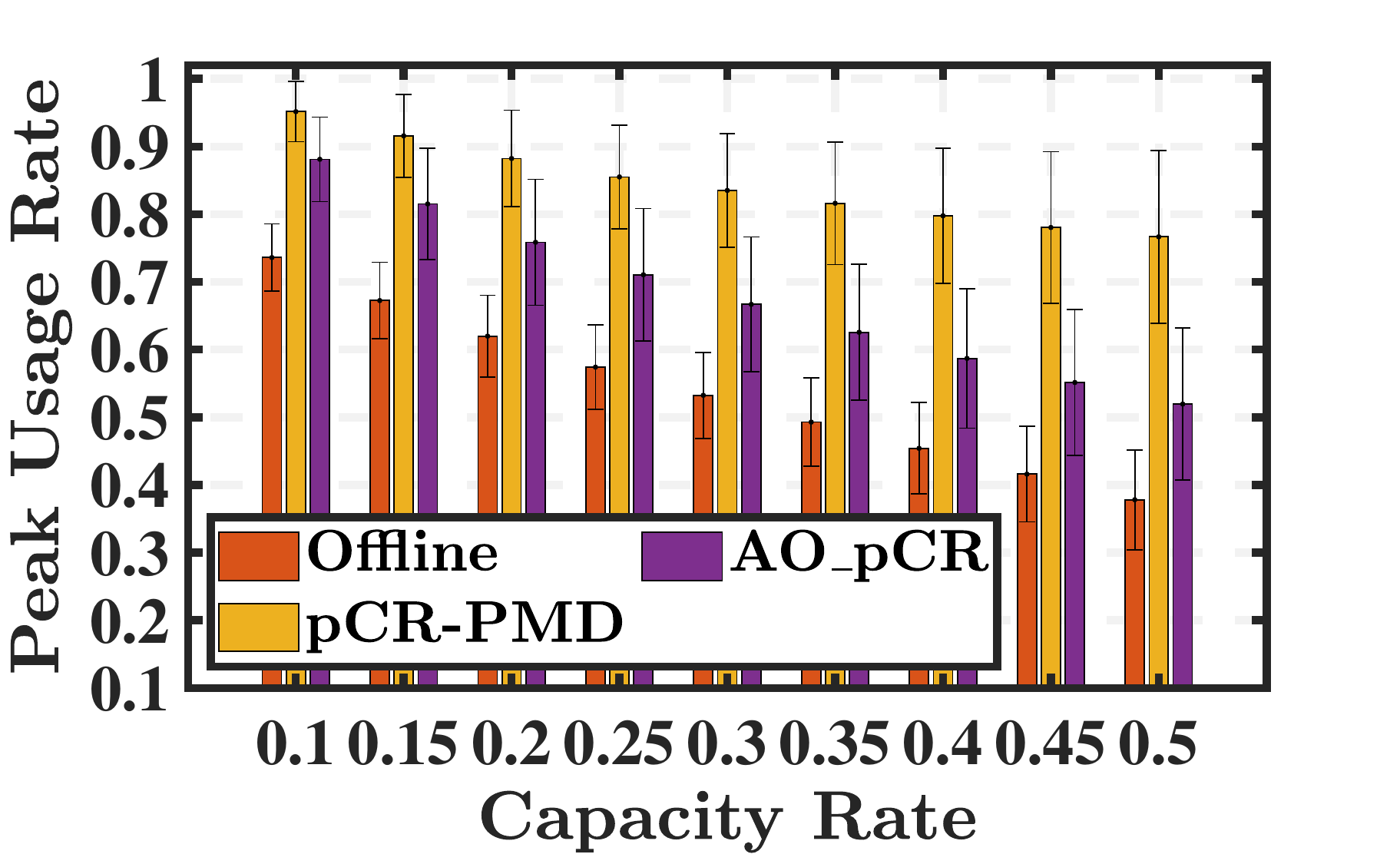}}
\caption{Peak usages under optimal offline solutions~(\emph{ref.}~Offline), \textsf{pCR-PMD}($\pi^*$)~(\emph{ref.}~pCR-PMD), and the anytime-optimal \textsf{pCR-PMD}~(\emph{ref.}~AO\_pCR).}
  \label{fig: comparisons_ada}
\end{figure}

In Section~\ref{Sec_Adaptive}, we have shown that the anytime-optimal~\textsf{pCR-PMD} has the same CR with~\textsf{pCR-PMD}($\pi^*$), while the former in principle presents an adaptive average-case performance. We herein observe from our real-world simulations that the anytime-optimal~\textsf{pCR-PMD}~(\emph{ref.}~AO\_pCR) attains better empirical performance than~\textsf{pCR-PMD}($\pi^*$)~(\emph{ref.}~pCR-PMD) under a wide range of storage capacities, as shown in Fig.~\ref{fig: comparisons_ada}. The peak usage rate herein refers to the ratio between the peak purchase after storage discharging and the original peak demand. We normalize the storage capacity and the capacity rate refers to the ratio of the storage capacity and the average daily energy consumption~($13,083$~kWh). We delineate the average peak usage rates and the corresponding standard deviations in Fig.~\ref{fig: comparisons_ada}. Specifically, we observe that, on average, the peak usage reduction under the anytime-optimal~\textsf{pCR-PMD} is more than twice that under~\textsf{pCR-PMD}($\pi^*$), while the peak usage rate under the anytime-optimal~\textsf{pCR-PMD} decreases faster than that under~\textsf{pCR-PMD}($\pi^*$) as the storage capacity increases. Moreover, we define the empirical performance ratio of an algorithm as the ratio between the average peak usage under the algorithm and that under the optimal offline solution. In the typical setup from real-world traces, we list the best CRs~(namely~$\pi^*$), the empirical performance ratios of~\textsf{pCR-PMD}($\pi^*$), and the empirical performance ratios of the anytime-optimal~\textsf{pCR-PMD} under different storage capacities in Table~\ref{tab: cr_comp}. As we can see, the anytime-optimal~\textsf{pCR-PMD} attains smaller empirical performance ratios compared with~\textsf{pCR-PMD}($\pi^*$). These observations substantiate the significance of absorbing real-time information and the advantage of the anytime-optimal~\textsf{pCR-PMD} algorithm over its non-adaptive counterpart~--~\textsf{pCR-PMD}($\pi^*$).

\begin{table}[t!]
\centering
\vspace{-.5\baselineskip}
  \caption{The online-to-offline ratios of peak usage. }
  \label{tab: cr_comp}
  \begin{tabular}{r|c|c|c|c|c}
    \toprule
    \hline
    Capacity Rate&0.10&0.20&0.30&0.40&0.50\\
    \hline
    $\pi^*$&1.3732   &  1.4621 &  1.6031   &  1.7953   & 2.0788\\
    pCR-PMD& 1.2909    &  1.4216    &  1.5669  &  1.7543    &  2.0244\\
    AO$\_$pCR& 1.1960  &  1.2236   &  1.2514    &  1.2912    &  1.3736\\
  \hline
  \bottomrule
\end{tabular}
\end{table}





We also observe from Table~\ref{tab: cr_comp} that the empirical performance ratios of the anytime-optimal~\textsf{pCR-PMD} are much smaller than the corresponding best possible CRs. That is because the worst cases regarding the optimal CRs rarely happen. In contrast with~\textsf{pCR-PMD}($\pi^*$) which ignores such non-occurrences and keeps maintaining the optimal CR~$\pi^*$, the anytime-optimal~\textsf{pCR-PMD} adaptively identifies the changing prices of future uncertainty and progressively pursues the anytime-optimal CRs across time instead. Moreover, we see that our anytime-optimal algorithm without knowing any future inputs achieves up to~$77\%$ of the peak usage reduction under the optimal offline outcomes where we have complete future knowledge. This fact means that our anytime-optimal algorithm achieves impressive empirical performance, in addition to the guaranteed best worst-case performance. The efficiency of our anytime-optimal~\textsf{pCR-PMD} will be further articulated by comparison with conceivable alternatives.

\subsubsection{Comparison with Baseline Alternatives}
We compare our anytime-optimal~\textsf{pCR-PMD} with two classes of conceivable alternatives: threshold-based and RHC algorithm\footnote{Note that the alternative algorithms cannot distinguish the two objectives, namely the peak minimization in this work and the peak-reduction maximization in~\cite{mo2021eEnergy}. In contrast, the algorithmic approach in this work differentiates the two complementary objectives. Thus, the algorithms proposed in this work and~\cite{mo2021eEnergy} are different, as illustrated in Appendix~\ref{app_expdif}.}.

    $\mbox{\ensuremath{\bullet}}$ The first class consists of four threshold-based algorithms. Specifically, two demand thresholds are respectively the average optimal offline peak purchase over the recorded days~(\emph{ref.}~THR\_avg) and the mean of net demand bounds, i.e.,~$(\underline{d}+\overline{d})/2$~(\emph{ref.}~THR\_half). The resulting two algorithms discharge the storage such that the demand reduces to the corresponding thresholds or the storage runs out. The remaining two algorithms~(\emph{ref.}~Eql\_Dis and~Eql\_Per) respectively maintain the discharge amount~$\delta_t$ as~$c/T$ and the discharging ratio~$\delta_t/d_t$ as the storage capacity rate.

    $\mbox{\ensuremath{\bullet}}$ The second class refers to three RHC algorithms with a quarter of the period as the look-ahead window size. The algorithms are different in their views on future demands, respectively assuming each future demand as the upper bound~(\emph{ref.}~RHC\_ub), the lower bound~(\emph{ref.}~RHC\_lb), and the mean of the two bounds~(\emph{ref.}~RHC\_half). Note that if RHC\_ub reduces the window size to one, the algorithm can be treated an adaptive robust one concerning the absolute worst-case performance.

\begin{figure*}
\centering
  \begin{subfigure}[t]{0.46\textwidth}
  \centerline{\includegraphics[width=1.0\columnwidth]{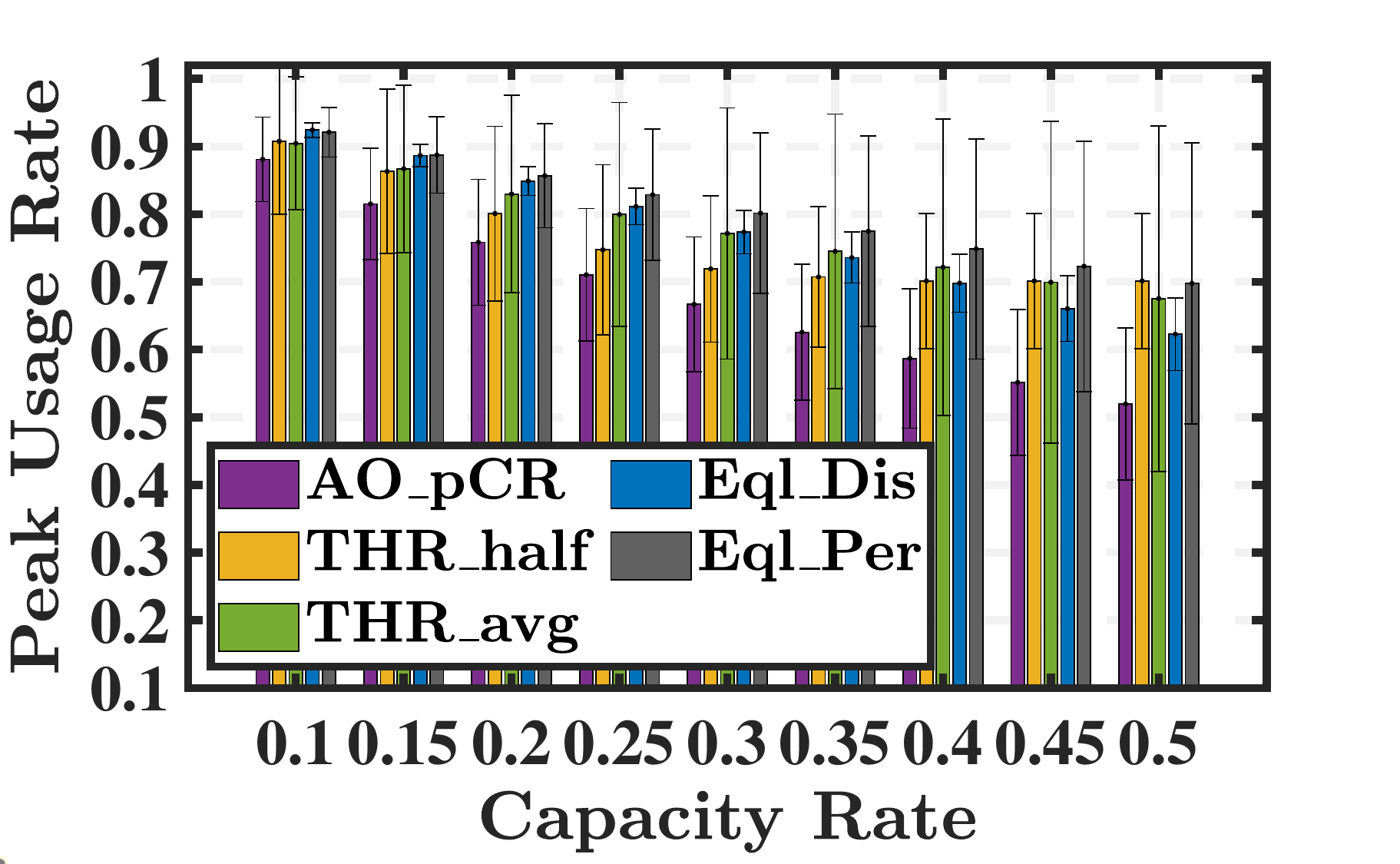}}
  \caption{}\label{fig: comparisons_alter}
  \end{subfigure}
  \begin{subfigure}[t]{0.46\textwidth}
  \centerline{\includegraphics[width=1.0\columnwidth]{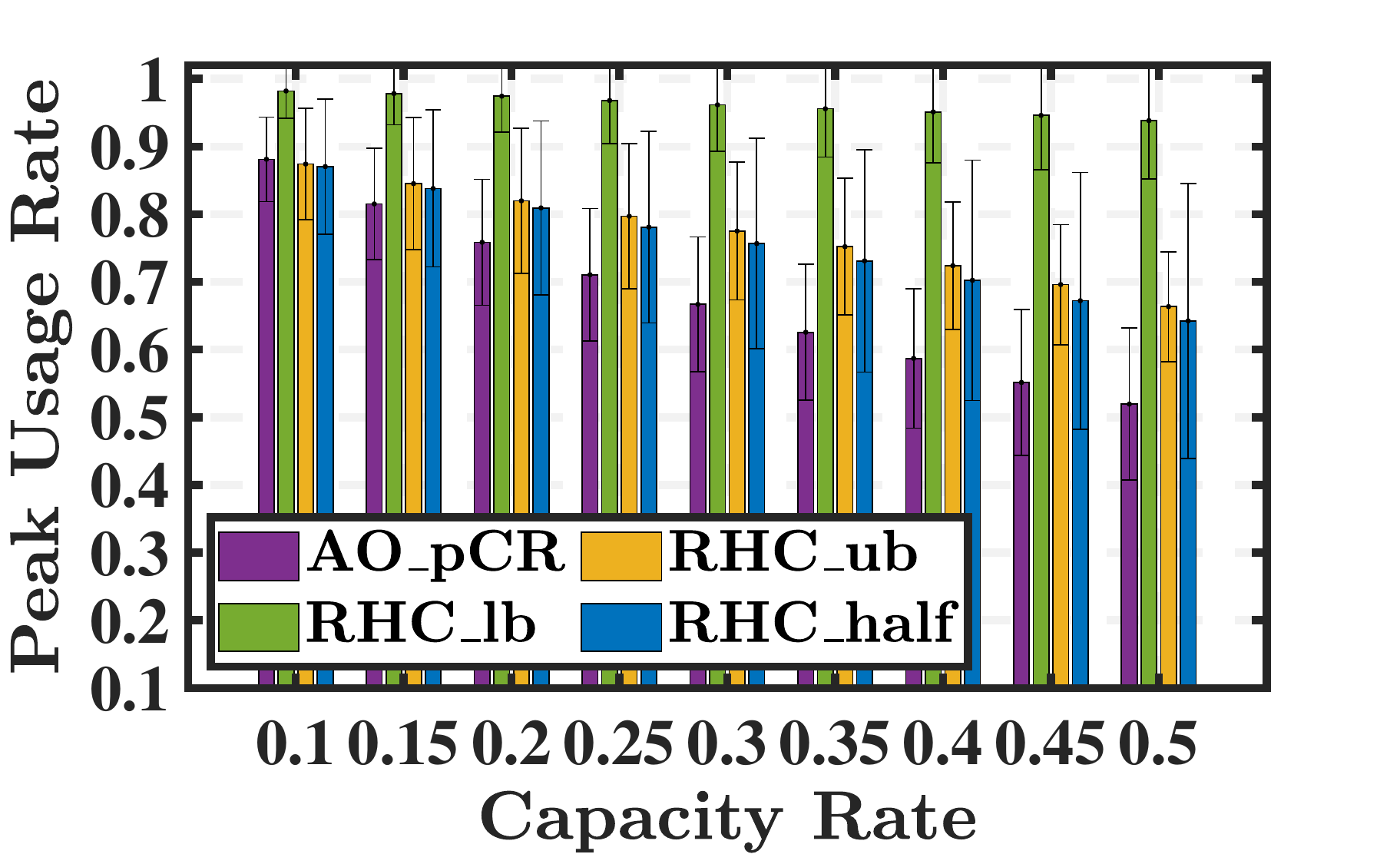}}
  \caption{}\label{fig: comparisons_rhc}
  \end{subfigure}
  \caption{
  (a)~Peak usages under four threshold-based algorithms and the anytime-optimal \textsf{pCR-PMD}. (b)~Peak usages under three RHC algorithms and the anytime-optimal \textsf{pCR-PMD}.
  }
\end{figure*}

 We observe from Fig.~\ref{fig: comparisons_alter} and Fig.~\ref{fig: comparisons_rhc} that our algorithm attains much better empirical performance than the threshold and RHC algorithms under a wide range of storage capacities. Particularly, when the storage capacity rate is~$30\%$, the anytime-optimal~\textsf{pCR-PMD} can improve the average peak usage reduction by at least~$19\%$ compared to the alternatives. Moreover, as the storage capacity increases, the average peak usage rate of our anytime-optimal~\textsf{pCR-PMD} decreases much faster than that of any baseline algorithm.

\begin{figure}[t]
    \centerline{\includegraphics[width=0.6\columnwidth]{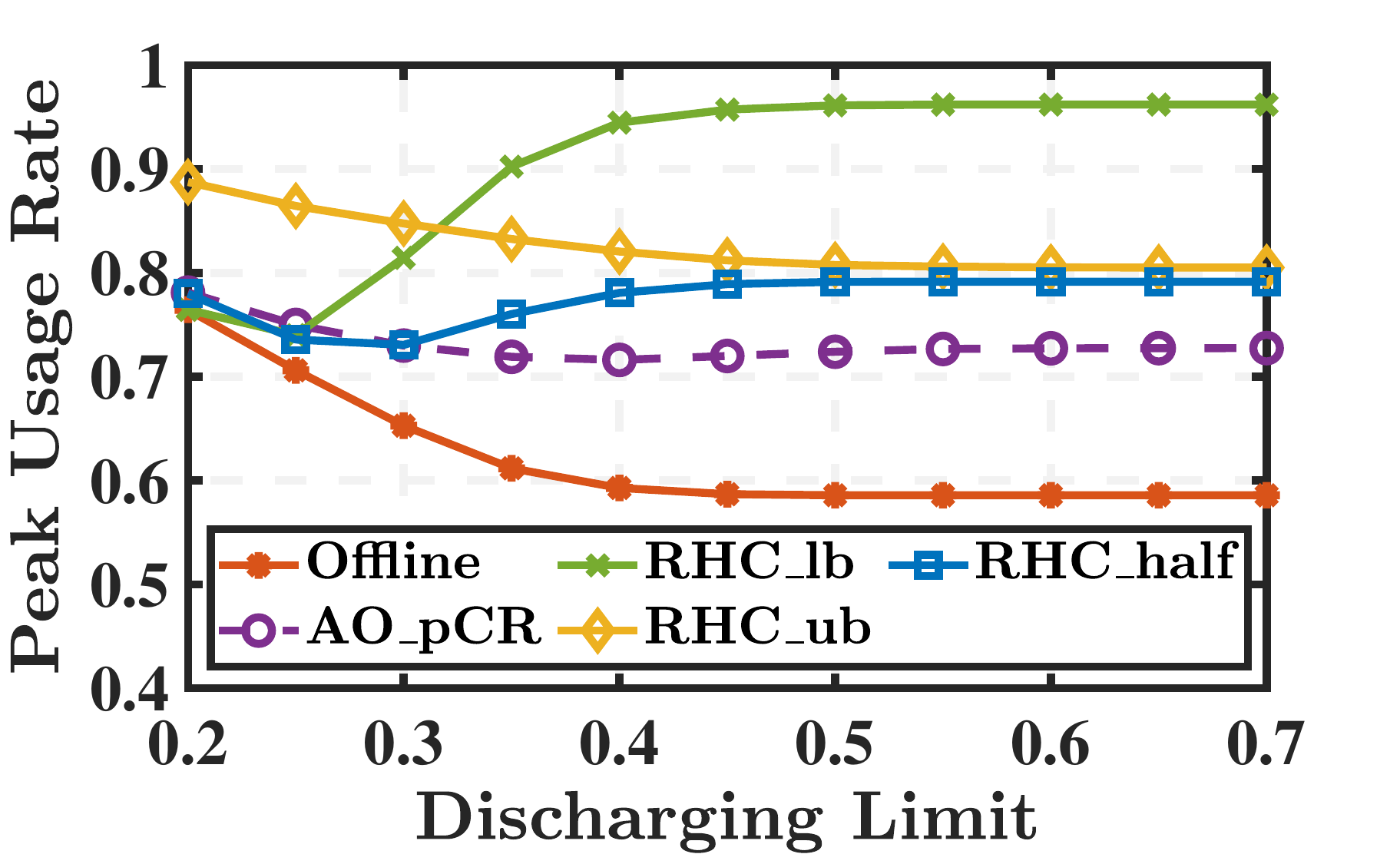}}
\caption{Impact of the maximum discharge rate.}
  \label{fig: comparisons_rate}
\end{figure}

\subsubsection{Impact of Discharging Limit}
We show the impact of the maximum discharge amount per time slot ($\bar{\delta}$) in Fig.~\ref{fig: comparisons_rate}, where the capacity is fixed as~$3,000$~kWh, almost~$23\%$ of the average daily demand. The discharging limit~($\bar{\delta}/\overline{d}$) refers to the ratio between the maximum discharge amount and the upper bound of the net demand in a time slot. Increasing the limit will improve the optimal offline outcome. However, it may worsen the outcomes of certain online algorithms, e.g. RHC\_lb and RHC\_half. The peak usage rate of our anytime-optimal~\textsf{pCR-PMD} first decreases and then slightly increases as we raise the discharging limit. Since the turning point is~$ 0.434\approx \underline{d}/\overline{d}$, a possible reason for the non-monotonicity may be that the maximum discharge amount exceeds the lower bound of the net demand when the assumption~(\ref{assumpc}) holds. Also, we conclude from Fig.~\ref{fig: comparisons_rate} that the discharging rate need not to be very large, because it has less impact on the anytime-optimal~\textsf{pCR-PMD} and will no longer reduce the optimal offline usage after exceeding a certain value.

\begin{figure}[t]
     \centerline{\includegraphics[width=0.56\columnwidth]{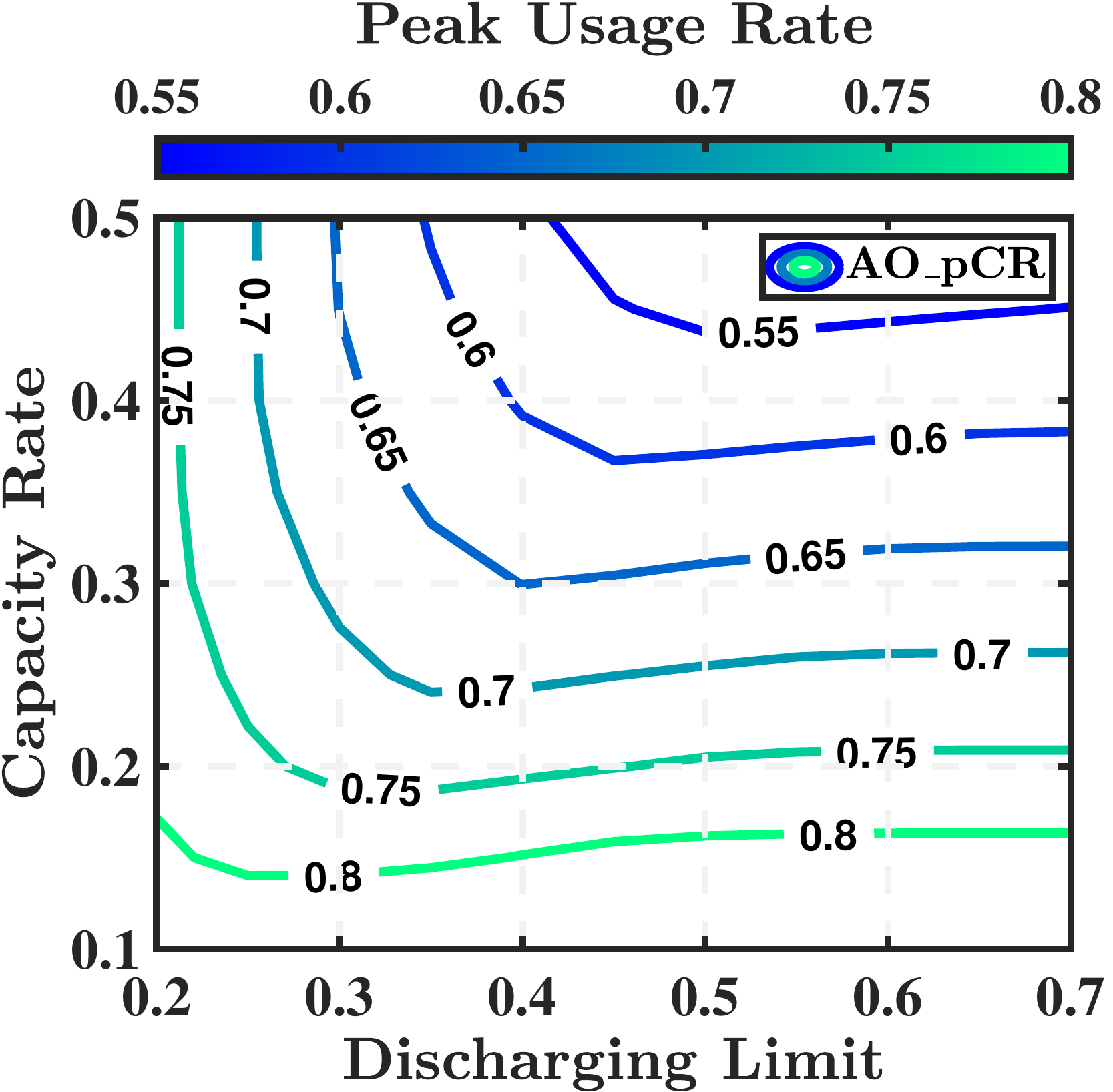}}
\caption{\rev{Achieving specific average peak usages by different combinations of energy storage capacity and discharging limit.}}
  \label{fig: comparisons_rate_rate}
\end{figure}

\rev{

\subsubsection{Achieving Specific Average Peak Usages by Different Combinations of Storage Capacity and Discharging Limit} Using our anytime-optimal~\textsf{pCR-PMD} algorithm, we evaluate the peak usages achieved by energy storage systems with different parameters, specifically the capacity and the discharging limit, under the online setting. Such evaluation can provide insights for large-load users to better plan their energy storage systems based on targeting peak usages, with demand uncertainty taken into account.

We plot a contour in Fig.~\ref{fig: comparisons_rate_rate} to illustrate how the capacity and the discharging limit jointly affect the peak usages achieved by the anytime-optimal~\textsf{pCR-PMD} algorithm under the online setting specified in Sec.~\ref{ssec_sim_setup}. A contour line in Fig.~\ref{fig: comparisons_rate_rate} specifies the capacity-limit combinations with which the anytime-optimal~\textsf{pCR-PMD} algorithm achieves the same average peak usage rate (the value is shown on the line). We observe from Fig.~\ref{fig: comparisons_rate_rate} that it is more effective in reducing peak usage to increase the discharging limit for the storage systems with a large capacity than otherwise. For example, when the discharging limit is 0.2, increasing the capacity rate from 0.2 to 0.4 merely reduces the peak usage rate by less than 0.05. In contrast, when the discharging limit is 0.4, the same increment in capacity rate will reduce the peak usage rate by around 0.15. Similarly, it becomes less effective to invest in extra energy storage to reduce peak usages for energy storage systems with a small discharging limit.
}

\rev{
\subsubsection{Monthly Peak Demand} \label{sim: monthly}

So far, we have focused on the effect of our algorithms in reducing daily peak demands. How can we adapt the algorithms to reduce the demand charge settled on a monthly basis? We apply the approach discussed in Section~\ref{DaytoMonth} to sequentially shave the peak usage over the on-peak period of each day and adjust the anytime-optimal ratios~$\pi^*_t$ taking into account the existing peak usages over the previous days of the same month. Theoretically, this practice should improve the effect of decreasing the monthly peak demand because we avoid the situation where the operator excessively discharges the energy storage such that the electricity procurement of a slot in a day is less than the peak usage over the previous days. Thus, the operator reserves more energy storage for shaving the peak usages that appear in later time slots in the on-peak period of the day. Recall that we are using 3-month traces; then, we will separately conduct numerical simulations under five different storage capacities for each of the three months.

We show in Table~\ref{tab: month_pdf} how much we further cut the monthly peak demand as compared to the case where we reduce the daily peak usage independently. For example, when the storage capacity rate is~$30\%$, the peak demands of month~$2$ after applying our adaptive algorithms with and without considering existing peak usages are~$726.04$~kWh and~$865.34$~kWh, respectively. Hence, the monthly peak demand decreases by extra~$16.10\%(=(865.34-726.04)/865.34)$. We observe the decreases in monthly peak demands for all fifteen instances, consistent with our theoretical analysis. This observation also clarifies the importance of exploiting the sequentially revealed information (peak usages in previous days) for online decision-making. Moreover, we see that the decreasing effect in Month~$3$ is less significant than in the other two months. A possible reason is that the peak demand of Month~$3$ appears in an earlier time slot in the on-peak period of the corresponding day. In this case, it is less likely to cause excessive discharges, even if the operator ignores the peak usages in previous days.}

\begin{table}[t!]
\centering
\vspace{-.5\baselineskip}
  \caption{\rev{Further Reduction in Monthly Peak Demand.}}
  \label{tab: month_pdf}
  \rev{
  \begin{tabular}{r|c|c|c|c|c}
    \toprule
    \hline
    \diagbox[width=12em, height=3em]{Month}{(\%)}{Capacity Rate}&0.10&0.20&0.30&0.40&0.50\\
    \hline
    Month~1&9.30  &  10.87 &  12.32 &15.50&20.66 \\
    Month~2&8.34  &  13.35 &  16.10 &15.06&15.98 \\
    Month~3&8.11  &  5.10 &  6.53 &4.17&7.69 \\
  \hline
  \bottomrule
\end{tabular}}
\end{table}

\section{Conclusions and Future Work}\label{Sec_Conclusion}

We study online peak-demand minimization under energy storage constraints and develop an optimal online algorithm. It achieves the best possible CR among all deterministic and randomized online algorithms. We show that the optimal CR~$\pi^*$ can be computed by solving a linear number of linear-fractional programs, incurring only polynomial time complexity. To our best knowledge, these are the first (and optimal) results for this theoretically challenging yet practically relevant problem. Furthermore, we generalize our approach to design an anytime-optimal algorithm to retain the optimal worst-case performance and achieve adaptive average-case performance. The idea is to adaptively prune the input space based on the inputs observed so far and adjust the online decisions to achieve the anytime-optimal CR concerning the residual uncertainty. Finally, we demonstrate the empirical efficiency of our algorithms by simulations based on real-world traces.

A compelling future direction is to extend the study to energy storage systems with both charging and discharging during operation. It is also interesting to design online algorithms with good performance guarantees for other cost functions 
e.g., the total volume and peak demand charge, with storage degradation effects taken into account.
\bibliographystyle{ACM-Reference-Format}


\appendix

\section{Proof of Proposition~\ref{prop_rand}}\label{app_rand}
  Without loss of generality, we can assume that Algorithm~$\mathfrak{A}$ is a mixed algorithm for {PMD}, which is a probability distribution~$\{\omega(\mathfrak{C})\}$ over a collection of deterministic online algorithms:~$\mathfrak{C}\in\mathcal{C}$. Given an arbitrary demand profile~$\bm d$, the expected peak purchased demand of Algorithm~$\mathfrak{A}$ is~$$\mathbf{E}[v^\mathfrak{A}(\bm d)]=\int_\mathcal{C}\max_{t\in[T]} (d_t-\delta_t^\mathfrak{C})d\omega(\mathfrak{C}),$$ where~$\delta_t^\mathfrak{C}$ is the discharging quantify under Algorithm~$\mathfrak{C}\in \mathcal{C}$ and the demand profile~$\bm d$. With respect to~$\bm d$, we consider a deterministic algorithm~$\mathfrak{B}$, under which the discharge amount of the~$t$th time slot is~$\int_{\mathcal{C}}\delta_t^\mathfrak{C}d\omega(\mathfrak{C})$, for all~$t\in [T]$. Since~${\bm \delta}^\mathfrak{C}$ is a feasible solution to~{PMD} for all~$\mathfrak{C}\in \mathcal{C}$, it follows that Algorithm~$\mathfrak{B}$ also generates a feasible solution to PMD. Then, the peak purchased demand under Algorithm~$\mathfrak{B}$ and the input sequence~$\bm d$ is
  \begin{equation*}
  \begin{split}
      v^\mathfrak{B}(\bm d)&=\max_{t\in T} \left(d_t-\int_{\mathcal{C}}\delta_t^\mathfrak{C}d\omega(\mathfrak{C})\right)\\
      &= \max_{t\in T} \left(\int_{\mathcal{C}}\left(d_t-\delta_t^\mathfrak{C}\right)d\omega(\mathfrak{C})\right)\\
      &\leq \int_{\mathcal{C}}\max_{t\in[T]}\left(d_t-\delta_t^\mathfrak{C}\right)d\omega(\mathfrak{C})=\mathbf{E}[v^\mathfrak{A}(\bm d)],
  \end{split}
  \end{equation*}
  where the second equality is due to the fact~$\int_{\mathcal{C}}1d\omega(\mathfrak{C})=1$ and the last inequality is due to the convexity of the~$\max$ function and Jensen's inequality. Moreover, by the definition of CR, we have
  $$CR_{\mathfrak{B}}=\max_{\bm d\in \mathcal{D}}\frac{v_{\mathfrak{B}}(\bm d)}{v(\bm d)}\leq \max_{\bm d\in \mathcal{D}}\frac{\mathbf{E}[v_{\mathfrak{A}}(\bm d)]}{v(\bm d)} =CR_{\mathfrak{A}}.$$
  Thus, for any randomized online algorithm~$\mathfrak{A}$ with its CR as~$CR_{\mathfrak{A}}$, we can find a deterministic online algorithm attaining a CR no larger than~$CR_{\mathfrak{A}}$.

\section{Proof of Proposition~\ref{propfeasibility}}\label{app_feasibility}
  The necessity follows from the definition. Now, we show the sufficiency. Since~$\Phi(\pi)\leq c$, the inventory constraint is satisfied by~$\bm \delta(\pi,\bm d)$ for all~$\bm d\in \mathcal{D}$. Moreover, we see that~
  \begin{displaymath}
    d_t-v(\bm d^t)\leq \min\{\bar{\delta},d_t\}\text{, for all }\bm d\text{ and }t\in[T].
  \end{displaymath} It follows that~$[d_t - \pi v(\bm d^t)]^+ \leq \min\{\bar{\delta},d_t\}\text{, for all }t\in [T],$
  by the facts~$\pi\geq 1$, $\min\{\bar{\delta},d_t\}\geq 0$, and~$v(\bm d^t)>0$. This completes the proof.

\section{Proof of Lemma~\ref{lemdeinc}}\label{app_deinc}
  First, we observe that for any input sequence~$\bm d$, the following function indexed by~$t\in[T]$ is strictly decreasing over~$\pi$ when its function value is positive:~$\delta_t(\pi,\bm d)=[d_t - \pi v(\bm d^t)]^+.$ Then, for any fixed input sequence~$\bm d$, we summarize the above functions over~$t\in[T]$ and obtain another nonincreasing function over~$\pi$:~$\sum_{t=1}^{T}\delta_t(\pi,\bm d),$ which is also strictly decreasing when its function value is positive. Thus, we can conclude that~$\Phi(\pi)$ is strictly decreasing before it attains zero.

\section{Proof of Theorem~\ref{thmoptratio}}\label{app_optratio}

  The uniqueness of the solution to~$\Phi(\pi)=c$ follows from Lemma~\ref{lemdeinc}. Also, by Proposition~\ref{prop_rand}, it suffices to consider deterministic online algorithms for PMD. Let~${\bm \delta}(\mathfrak{A},\bm d)$ be an output sequence of any other deterministic online algorithm~$\mathfrak{A}$ for {PMD} under the input sequence~$\bm d$. Let~$\hat{\bm d}$ be a worst-case input sequence of~\textsf{pCR-PMD}($\pi^*$), namely,~$\sum_{t=1}^{T}\delta_t(\pi^*,\hat{\bm d})=c$. Then, we shall construct an input sequence over which the algorithm~$\mathfrak{A}$ cannot achieve an offline-to-online ratio of peak usage which is strictly smaller than~$\pi^*$.

  Let~$\bm d$ be an input sequence with~$\bm d\in \mathcal{D}$ and~$d_1=\hat{d}_1$. If~${\delta}_1(\mathfrak{A},\bm d)<\delta_1(\pi^*,\bm d)$, then we have
  \begin{equation*}
    \max_{t\in[T]} (d_t-{\delta}_t(\mathfrak{A},\bm d))\geq d_1-{\delta}_1(\mathfrak{A},\bm d)>d_1-\delta_1(\pi^*,\bm d).
  \end{equation*}
  It follows from~$\delta_1(\pi^*,\bm d)>0$ that~$d_1-\delta_1(\pi^*,\bm d)=\pi^* v(\bm d^1)$. Thus, if it happens that~$\bm d=\hat{\bm d}^1$, then the peak usage under the online algorithm~$\mathfrak{A}$ is strictly larger than~$\pi^*$ times the optimal offline objective value of {PMD}. This implies that the CR of the algorithm~$\mathfrak{A}$ is strictly bigger than~$\pi^*$.

  If~${\delta}_1(\mathfrak{A},\bm d)\geq \delta_1(\pi^*,\bm d)$, then we continue by sequentially presenting~$d_t=\hat{d}_t$, for~$t=2,3,\ldots,T$. If~${\delta}_t(\mathfrak{A},\bm d)=\delta_t(\pi^*,\bm d)$ for all~$t\in [T]$, let~$\hat{t}$ be the largest index with~\mbox{$\delta_t(\pi^*,\bm d)>0$}. With the input sequence~$\bm d=[d_1~d_2~\cdots~d_{\hat{t}}~\underline{d}~\cdots,\underline{d}]$, we can show that the online-to-offline ratio of the peak usage under~$\mathfrak{A}$ is~$\pi^*$. Otherwise, by the feasibility of ${\bm \delta}(\mathfrak{A},\bm d)$, there exists an index~$\hat{t}$ such that~${\delta}_{\hat{t}}(\mathfrak{A},\bm d)< \delta_{\hat{t}}(\pi^*,\bm d)$. Moreover, with~$\bm d=[d_1~d_2~\cdots~d_{\hat{t}}~\underline{d}~\cdots,\underline{d}]$, we have
  \begin{equation*}
    \max_{t\in[T]} (d_t-{\delta}_t(\mathfrak{A},\bm d))\geq d_{\hat{t}}-{\delta}_{\hat{t}}(\mathfrak{A},\bm d)>d_{
    \hat{t}}-\delta_{\hat{t}}(\pi^*,\bm d).
  \end{equation*}
  It follows from~$\delta_{\hat{t}}(\pi^*,\bm d)>0$ that~$d_{\hat{t}}-\delta_{\hat{t}}(\pi^*,\bm d)=\pi^* v(\bm d)$. Thus, if it happens that~$\bm d=\hat{\bm d}^{\hat{t}}$, then the peak usage under the online algorithm~$\mathfrak{A}$ is strictly larger than~$\pi^*$ times the optimal offline objective value of {PMD}. This implies that the CR of the online algorithm~$\mathfrak{A}$ is bigger than~$\pi^*$.

  Thus, we have shown that no other deterministic online algorithm can attain a CR strictly smaller than~$\pi^*$. By Proposition~\ref{prop_rand} and the CR of~\textsf{pCR-PMD}($\pi^*$), we complete the proof.

\section{Proof of Lemma~\ref{lemparwc}}\label{app_parwc}

  By Theorem~\ref{thmoptratio}, we can find an input sequence~$\hat{\bm d}$ which satisfies~$\sum_{t=1}^{T}\delta(\pi^*,\hat{\bm d})=c$. If there is an index~$i$ such that~$\delta_t(\pi^*,\hat{\bm d})>0$ whenever~$t\in [i]$ and~$\delta_t(\pi^*,\hat{\bm d})=0$ whenever~$t>i$, then we complete the proof. Otherwise, we observe that there exists an index~$i$ such that
  \begin{equation*}
    \hat{d}_i-\pi v(\pi^*,\hat{\bm d}^i)\leq 0 \text{  and  } \hat{d}_{i+1}-\pi v(\pi^*,\hat{\bm d}^{i+1})>0.
  \end{equation*}
  It follows from~$v(\pi^*,\hat{\bm d}^i)\leq v(\pi^*,\hat{\bm d}^{i+1})$ that~$\hat{d}_i<\hat{d}_{i+1}$. Now, let us construct another input sequence~$\tilde{\bm d}$ from~$\hat{\bm d}$ by exchanging~$\hat{d}_i$ and~$\hat{d}_{i+1}$. Then, it holds that
  \begin{align*}
    &\tilde{d}_i-\pi v(\pi^*,\tilde{\bm d}^i)\geq \hat{d}_{i+1}-\pi v(\pi^*,\hat{\bm d}^{i+1})>0 \text{ and } \\
    &\tilde{d}_{i+1}-\pi v(\pi^*,\tilde{\bm d}^{i+1})\leq \hat{d}_i-\pi v(\pi^*,\hat{\bm d}^i)\leq 0.
  \end{align*}
  It follows that~$\delta_i(\pi^*,\tilde{\bm d})\geq \delta_{i+1}(\pi^*,\hat{\bm d})>0\text{ and }\delta_i(\pi^*,\hat{\bm d})=\delta_{i+1}(\pi^*,\tilde{\bm d})=0.$ Moreover, we see that~$\delta_t(\pi^*,\hat{\bm d})=\delta_t(\pi^*,\tilde{\bm d})$ for all~$t\in[T]$ and~$t\notin \{i, i+1\}$. It follows that
  \begin{equation*}
    c=\Phi(\pi^*)\geq \sum_{t=1}^{T}\delta_t(\pi^*,\tilde{\bm d})\geq \sum_{t=1}^{T}\delta_t(\pi^*,\hat{\bm d})=c,
  \end{equation*}
  which further implies that~$\sum_{t=1}^{T}\delta_t(\pi^*,\tilde{\bm d})=c$. Following similar arguments, by a series of exchanges of adjacent elements from~$\hat{\bm d}$, we can finally construct a worse-case input sequence~$\bm d$ with an index~$i$ such that~$\delta_t(\pi^*,\bm d)>0$ if~$t\in [i]$ and~$\delta_t(\pi^*,\bm d)=0$ otherwise.

\section{Differences between Online PMD and Online PRM} \label{app_expdif}

To see the difference concretely, let us consider the following system setting:
\begin{equation}
    c=630,\, T=10,\, \underline{d}=300,\, \overline{d}=600,\, \bar{\delta}=\infty.    \label{eq:PMD.vs.prm.setting}
\end{equation}
Under the offline setting where the demands~$\{d_t\}_{t=1}^{T}$ are given beforehand, the PMD and PRM problems share the same optimal solutions, i.e., their optimal discharging amounts are identical for each slot~$t\in[T]$, as suggested by Proposition~\ref{offopt} in this paper and~\cite{mo2021eEnergy}. In the following, we show that they are fundamentally different under the online setting, where the demands are revealed sequentially and decisions are irrevocable.

First, we show that the two problems admit different optimal CRs under the same setup in~\eqref{eq:PMD.vs.prm.setting}. Specifically, by the procedure in Sec.~\ref{ssec:opt_pi}, we obtain the optimal CR for the online PMD problem to be~$\pi^{*, PMD} = 1.32$, which is different  from the optimal CR for the online PRM problem, i.e., $\pi^{*, PRM}=2.73$, as computed by the method in~\cite{mo2021eEnergy}. The difference in the optimal CRs means that the same demand uncertainty has different impacts on the best achievable competitiveness for the two problems; thus they are \emph{structurally} different under the online setting.



Second, we show that the two online problems admit different optimal online algorithms. Specifically, we consider the optimal \textsf{pCR-PMD}($\pi^{*, PMD}$) algorithm for the PMD problem in this paper and the optimal \textsf{pCR-PRM}($\pi^{*, PRM}$) algorithm for the PRM problem in~\cite{mo2021eEnergy}. Table~\ref{tab: disam} shows that the two algorithms generate different online solutions, i.e., per-slot discharging amounts, for the following demand sequence:~$
    \hat{\bm d}=[379.5~411~411~442.5~442.5~600~600~600~600~600]'.
$ This is in contrast to the offline setting where the offline solutions to the two problems are identical for the same input.

Furthermore, we show that the \textsf{pCR-PMD}($\pi^{*, PMD}$) algorithm, optimal for the PMD problem, is not competitive at all for the PRM problem (i.e., the corresponding CR is unbounded). Specifically, we compute the performance ratios of the \textsf{pCR-PMD}($\pi^{*, PMD}$) algorithm regarding the PMD and PRM problems, respectively\footnote{\rev{Note that the performance ratios for an online minimization problem and an online maximization problem differ in definition. The ratio is online performance over offline performance for the former and is offline performance over online performance for the latter.}}, as follows:
\begin{align*}
    \frac{v_{\text{\textsf{pCR-PMD}}}(\hat{\bm d})}{v_{\textsf{offline}}(\hat{\bm d})} = \frac{600}{474} < \pi^{*, PMD}=1.32,
\text{ and }
    \frac{\max_{t\in[T]}\hat{d}_t-v_{\textsf{offline}}(\hat{\bm d})}{\max_{t\in[T]}\hat{d}_t-v_{\text{\textsf{pCR-PMD}}}(\hat{\bm d})} = \frac{600-474}{600-600}\;\;\; \mbox{(unbounded)}.
\end{align*}
These calculations show that (i) the \textsf{pCR-PMD}($\pi^{*, PMD}$) algorithm does not reduce the overall peak for this particular input sequence $\hat{\bm d}$, (ii) the zero peak reduction for $\hat{\bm d}$ is within the best possible competitiveness guarantee for the online PMD problem, and (iii) the zero peak reduction for $\hat{\bm d}$ results in an unbounded performance ratio for one input $\hat{\bm d}$ for the PRM problem (and thus an unbounded CR by definition).

\begin{table}[t!]
\centering
  \caption{{Online solutions of different algorithms (i.e., discharging amounts for each slot), given the input sequence~$\hat{\bm d}=[379.5~411~411~442.5~442.5~600~600~600~600~600]'$.}}
  \label{tab: disam}
  \rev{
  \begin{tabular}{r|c|c|c|c|c}
    \toprule
    \hline
    \diagbox[width=6.5em, height=3em]{\raisebox{1.5pt}{\hspace*{-0.65em}Algorithm}}{\raisebox{-3pt}{\hspace*{0.6em}Slot}}&1&2&3&4&5\\
    \hline
    \textsf{pCR-PMD}($\pi^{*, PMD}$)  & 56.10 & 72.94 & 58.28 & 70.97 & 52.16  \\
    \textsf{pCR-PRM}($\pi^{*, PRM}$) & 49.23 & 56.70 & 52.64 & 58.95 & 53.73 \\
  \hline
  \hline
    \diagbox[width=6.5em, height=3em]{\raisebox{1.5pt}{\hspace*{-0.65em}Algorithm}}{\raisebox{-3pt}{\hspace*{0.6em}Slot}}&6&7&8&9&10\\
    \hline
    \textsf{pCR-PMD}($\pi^{*, PMD}$)   & 147.47 & 98.95 & 57.36& 15.77 & 0 \\
    \textsf{pCR-PRM}($\pi^{*, PRM}$)  & 94.13 & 80.68 & 69.16 & 57.63 & 46.10 \\
  \hline
  \bottomrule
\end{tabular}}
\end{table}

Thus, the \textsf{pCR-PMD}($\pi^{*, PMD}$) algorithm that is optimal for the online PMD problem is NOT competitive for the online PRM problem. A similar calculation shows that the \textsf{pCR-PRM}($\pi^{*, PRM}$) algorithm that is optimal for the online PRM problem is not optimal for the online PMD problem.


In summary, the PMD and PRM problems are different under the online setting, admitting different optimal CRs and optimal competitive online algorithms. This observation, together with that the online optimizations of peak and peak reduction can be of independent interests to different parties in practice (similar to the $k$-max and $k$-min search problems in the literature), motivates our studies in this paper for online PMD and in~\cite{mo2021eEnergy} for online PRM.

\end{document}